\documentclass[]{aastex7}

\usepackage{amsmath}
\usepackage{xcolor}
\usepackage{cleveref}
\usepackage{siunitx}
\DeclareSIUnit{\gauss}{G}
\definecolor{darkgreen}{RGB}{0,127,0}

\usepackage{graphicx}
\DeclareGraphicsExtensions{.pdf,.png,.jpg}

\providecommand{\e}[1]{\ensuremath{\times 10^{#1}}}

\received{\today}
\revised{\today}
\accepted{\today}
\submitjournal{ApJ}

\shorttitle{3D Dynamics of Flare Reconnection}
\shortauthors{Dahlin et al.}

\begin{document}

\title{Determining the 3D Dynamics of Solar Flare Magnetic Reconnection}

\author[0000-0002-9493-4730]{Joel T. Dahlin}

\affiliation{Heliophysics Science Division, NASA Goddard Space Flight Center,
Greenbelt, MD, USA} 
\affiliation{Astronomy Department, University of Maryland, College Park, MD 20742, USA}
\email{joeltdahlin@gmail.com, joel.t.dahlin@nasa.gov}

\author{Spiro K. Antiochos}
\affiliation{Department of Climate and Space Sciences and Engineering, University of Michigan, Ann Arbor,
MI, USA} 
\email{spiro.antiochos@gmail.com}

\author{Peter F. Wyper}
\affiliation{Department of Mathematical Sciences, Durham University, Durham DH1 3LE, UK }
\email{peter.f.wyper@durham.ac.uk}

\author{Jiong Qiu}
\affiliation{Department of Physics, Montana State University, Bozeman, MT, USA}
\email{qiu@montana.edu}

\author{C. Richard DeVore}
\affiliation{Retired Independent Consultant}
\affiliation{Heliophysics Science Division, NASA Goddard Space Flight Center,
Greenbelt, MD, USA} 
\email{c.richard.devore@nasa.gov}

\correspondingauthor{Joel T. Dahlin}
\email{joeltdahlin@gmail.com, joel.t.dahlin@nasa.gov}

\begin{abstract}
Solar flares are major space weather events that result from the explosive conversion of stored magnetic energy into bulk motion, plasma heating, and particle acceleration. While the standard flare model has proven highly successful in explaining key morphological features of flare observations, many aspects of the energy release are not yet understood. In particular, the turbulent three-dimensional structure of the flare current sheet is thought to play an important role in fast reconnection, particle acceleration, and bursty dynamics. Although direct diagnosis of the magnetic field dynamics in the corona remains highly challenging, rich information may be gleaned from flare ribbons, which represent the chromospheric imprints of reconnection in the corona. Intriguingly, recent solar imaging observations have revealed a diversity of fine structure in flare ribbons that hints at corresponding complexity in the reconnection region. We present high-resolution three-dimensional MHD simulations of an eruptive flare and describe our efforts to interpret fine-scale ribbon features in terms of the current sheet dynamics. In our model, the current sheet is characterized by many coherent magnetic structures known as plasmoids. We derive a model analogue for ribbons by generating a time series of field-line length maps ($L$-maps) and identifying abrupt shortenings as flare reconnection events. We thereby demonstrate that plasmoids imprint transient `spirals’ along the analogue of the ribbon front, with a morphology consistent with observed fine structure. We discuss the implications of these results for interpreting SolO, IRIS, and DKIST observations of explosive flare energy release.
\end{abstract}

\keywords{}

\section{Introduction} \label{sec:intro}

Solar flares are explosive events in the Sun's corona that release as much as $10^{32}$ ergs in a matter of minutes. The vast majority of this energy is ultimately emitted across a broad range of the electromagnetic spectrum, illuminating a rich variety of phenomena involved in the energy release process. Among the most striking features of flares is their propensity to generate nonthermal particle populations, whose signatures range from radio emission triggered by electron beams to gamma rays revealing the presence of relativistic ions.
\par
The primary observed morphological features of solar flares, the bright chromospheric ribbons and intense X-Ray coronal loops, are widely believed to result from magnetic reconnection in the solar corona, as described in the so-called `standard model' for two-ribbon solar flares \citep[or CSHKP, after its originators:][]{carmichael64a,sturrock66a,hirayama74a,kopp76a}. While this model convincingly captured the primary observed properties of flares, it does not address secondary dynamics observed during the energy release processes. For example, there is clear observational evidence for turbulent dynamics \citep{kontar17a}, shocks \citep{chen15a}, coherent structures \citep{kumar13a,takasao16a,kumar19a}, bursty energy release \citep{aschwanden87a,nakariakov09a}, and nonthermal particle acceleration \citep{lin76a}, none of which is directly addressed by the standard model.
\par
Furthermore, while the standard model is 2D, advances in solar instrumentation have enabled observation of the detailed 3D structure of the flare energy release process. Many of important large-scale morphological features have been addressed via extensions to the standard model that accounts for the three-dimensional topology of the flux rope \citep[e.g., ribbon hooks encircling flux rope footpoints and the strong-to-weak shear transition of flare loops][]{janvier14a,aulanier12a,dahlin22a} and incorporates three-dimensional generalizations of reconnection \citep[e.g., quasi-separatrix layers and hyperbolic flux tubes and slipping/slip-running reconnection][]{demoulin96a,aulanier05a,aulanier06a,dudik14a}. However, these extensions do not address the nature of the fine-scale structure such as Supra-Arcade Downflows (SADs), dark finger-like downflows moving toward the flare arcade that reveal a highly structured reconnection outflow \citep{mckenzie00a,savage12a}. Coherent structures in the flare current sheet such as `blobs' and bidirectional outflows\citep{takasao12a,liu13a_w,kumar13a,kumar19a,kumar23a,kumar25a,cheng18a,yu20a,yan22a,kumar25a} are also indicative of bursty 3-D structures localized in space and time. Fine-scale features have also been identified in many observations of the flare ribbons which are interpreted to be the footprints of the coronal reconnection, and hence indirect proxies for those dynamics. Fine scale features in the ribbons has been interpreted as resulting from a variety of physical processes, including Kelvin-Helmholtz instabilities \citep{ofman11a,brannon15a}, slipping/slip-running reconnection\citep{dudik14a,dudik16a,lorincik24a}, or tearing instabilities \citep{brannon15a,french21a,wyper21a,faber25a}.
\par
The underlying physics of many of these features may be explained in terms of plasmoids, which are coherent magnetic structures generated during the reconnection process. Numerical modeling of reconnection indicates that plasmoids are a ubiquitous feature spanning plasma regimes from kinetic \citep{drake06b}, to magnetohydrodynamic \citep{loureiro07a,huang16a} scales, with a generic requirement that the dissipation scale be greatly separated from the global scale \citep[e.g.][]{shibata01a,ji11a}. In the solar corona, for example, the dissipation scale (e.g., the ion diffusion region) is of order meters, many orders of magnitude separated from typical flare dimensions $\gtrsim$ Mm. Plasmoids imply a bursty nature to the reconnection process, making them a promising explanation for fluctuating emission in flares. Furthermore, plasmoids may operate both as accelerators and traps for particles, and are therefore thought to be important for particle acceleration and transport \citep{drake06a,nishizuka13a,dahlin14a,dahlin20a,li15a_x}. 
\par
In order to understand the impact of plasmoids on particle acceleration and transport in observed flares, it is necessary to make quantitative measurement of their properties. While hot/dense propagating features in the flare current sheet\citep{takasao12a,liu13a_w,kumar13a,kumar19a,kumar23a,kumar25a,cheng18a,yu20a,yan22a} are suggestive of plasmoids, it is presently prohibitively difficult to measure the associated fine-scale magnetic structure, a challenge only further exacerbated by the necessity of disentangling unavoidable line-of-sight integration. These challenges are well-illustrated in recent Coronal Multi-channel Polarimeter \citep[CoMP][]{tomczyk08a} observations that indicated an absence of polarization near a flare current sheet, hinting at the presence of unresolved fine-scale structure \citep{french19a}. 
\par
While the three-dimensional structure of the flare current sheet is difficult to observe directly, detailed diagnostics of the magnetic field dynamics in the corona can be obtained from the so-called flare ribbons, rapid brightenings of the chromospheric footpoints of field lines reconnected in the solar corona. These brightenings sweep across the solar surface, providing information on the initiation and three-dimensional spread, allowing important properties of the reconnection, such as the rate at which flux is reconnected, to be measured accurately. Recent high resolution observations by 
NASA's Solar Dynamics Observatory (SDO) \citep{pesnell12a} Atmospheric Imaging Assembly (AIA) \citep{lemen12a} and Interface Region Imaging Spectrograph (IRIS) \citep{depontieu14a} have revealed intriguing fine structure at the smallest observable scales. For example, \textcite{ofman11a} found evidence of vortex-shaped features in EUV that the authors argued was evidence of Kelvin-Helmholtz instability. In another study, \textcite{li15a_t} used SDO/AIA and IRIS imaging of an eruptive X-class flare to investigate small-scale, quasi-periodic bright knots that propagated along a section of the ribbons and suggested this was evidence of slipping reconnection during the preflare stage \citep[similar features have been observed in a variety of other flares, e.g.,][]{li14a_t,li16a_t,li18a_t}. Slipping reconnection was similarly invoked to interpret IRIS observations of ribbon `squirming' \citep{dudik16a}, and remarkable super-Alfv\'enic apparent motions of bright flare kernels identified high-cadence IRIS observations have been argued to be evidence of \textit{slip-running} reconnection \citep{lorincik24a}. On the other hand, \textcite{brannon15a} identified oscillating, quasi-periodic structures in an IRIS observation of an M-class flare, which they argued was most likely due to a current sheet instability such as the tearing mode or Kelvin-Helmholtz instability. A string of blobs with length scales $\sim$ 140-200 km observed by the Swedish Solar Telescope (SST) were likewise suggested to be due to the tearing instability \citep{faber25a}. Goode Solar Telescope (GST) imaging has revealed individual flare kernels may be as small as ~80 km \citep{jing16a}, which represents the limits of present resolution. Further insights are likely by means of new short-exposure (0.04 s) EUV imaging by Solar Orbiter \citep{collier24a,chitta25a} and upcoming observations from the 4-meter Daniel K. Inouye Solar Telescope \citep[DKIST;][]{rimmele20a} which promise to further advance our understanding of the smallest-scale dynamics of flares.

High-resolution ribbon observations, therefore, represent a rich source of information on the time-varying and three-dimensional dynamics of flare reconnection. However, due to the indirect nature of the observations it is often unclear what particular mechanism is ultimately responsible for the fine structure, as evidenced by the competing interpretations discussed above. Recently, \textcite{wyper21a} (henceforth WP21) analyzed an analytical model of an eruptive flare in which they imposed small-scale magnetic twist to generate plasmoid-like structure in the current sheet. Plasmoid motions were approximated by varying the height of the imposed twist. They found that these structures distort the topological analogues of flare ribbons in characteristic `breaking-wave` and `spiral` patterns consistent with the IRIS observations, demonstrating a clear connection between the reconnection dynamics and flare ribbons. The qualitative features of the spirals/breaking waves were shown to carry information about the plasmoid topology (e.g., whether the structure is quasi-2D or instead oblique to the reconnection axis), and furthermore exhibited a chirality that matched that of the associated erupting CME flux rope. These both represent key predictions for testing the plasmoid model against observations. While the WP21 study presented important new insights on the link between plasmoids and ribbon structure, a key limitation was the absence of true time-dependent dynamics such as the self-consistent formation of plasmoids as part of the reconnection process. In this article we present results from full three-dimensional numerical modeling that captures both the complete evolution and structure of reconnection-generated plasmoids and determines their time-dependent imprint on the ribbon structure. Our results, therefore, can be compared directly to observations.
\par
We describe below our state-of-the-art high-resolution, 3D magnetohydrodynamics (MHD) model of an eruptive flare carefully designed to elucidate the signatures of 3D reconnection structure. We first present an overview of the eruptive flare model in \S\ref{sec:model}. We then discuss the structure of the flare current sheet (\S\ref{sec:plasmoids}) and show that the reconnection is characterized by many plasmoids that exhibit complex three-dimensional structure. We then discuss our method for identifying the analogues of flare ribbons in \S\ref{sec:ribbons}, and examine in detail the fine structure in these ribbons (\S\ref{sec:structure}), demonstrating a link to the corresponding plasmoid structures in the flare current sheet. We present observational predictions of plasmoid signatures in \S\ref{sec:observations} and discuss the implication of these results in \S\ref{sec:discussion}.

\section{Eruptive Flare Model} \label{sec:model}

First, we describe our numerical model for the eruptive flare. Our primary goal is to capture in detail the reconnection dynamics in the current sheet, and  in particular the formation and evolution of plasmoids as well as the topological connections between reconnection sites and the solar surface. To achieve this aim, we designed a high-resolution eruptive flare simulation that incorporated the formation and destabilization of the flare current sheet (i.e., included the build-up and onset phases). We chose for a magnetic configuration an idealized, highly symmetric profile to represent a generic two-ribbon morphology while introducing minimal extraneous details that would muddy the analysis. A detailed description of this configuration was presented in \textcite{dahlin22a}. In contrast to our previously presented simulation, we here employ a configuration with 6 levels of adaptive refinement, implying a factor of 4 less resolution than the simulations we presented in \textcite{dahlin22a}. This resolution is sufficient to produce many plasmoids, while allowing for detailed quantitative analysis of individual features.
\par
\begin{figure}[ht!]
\plotone{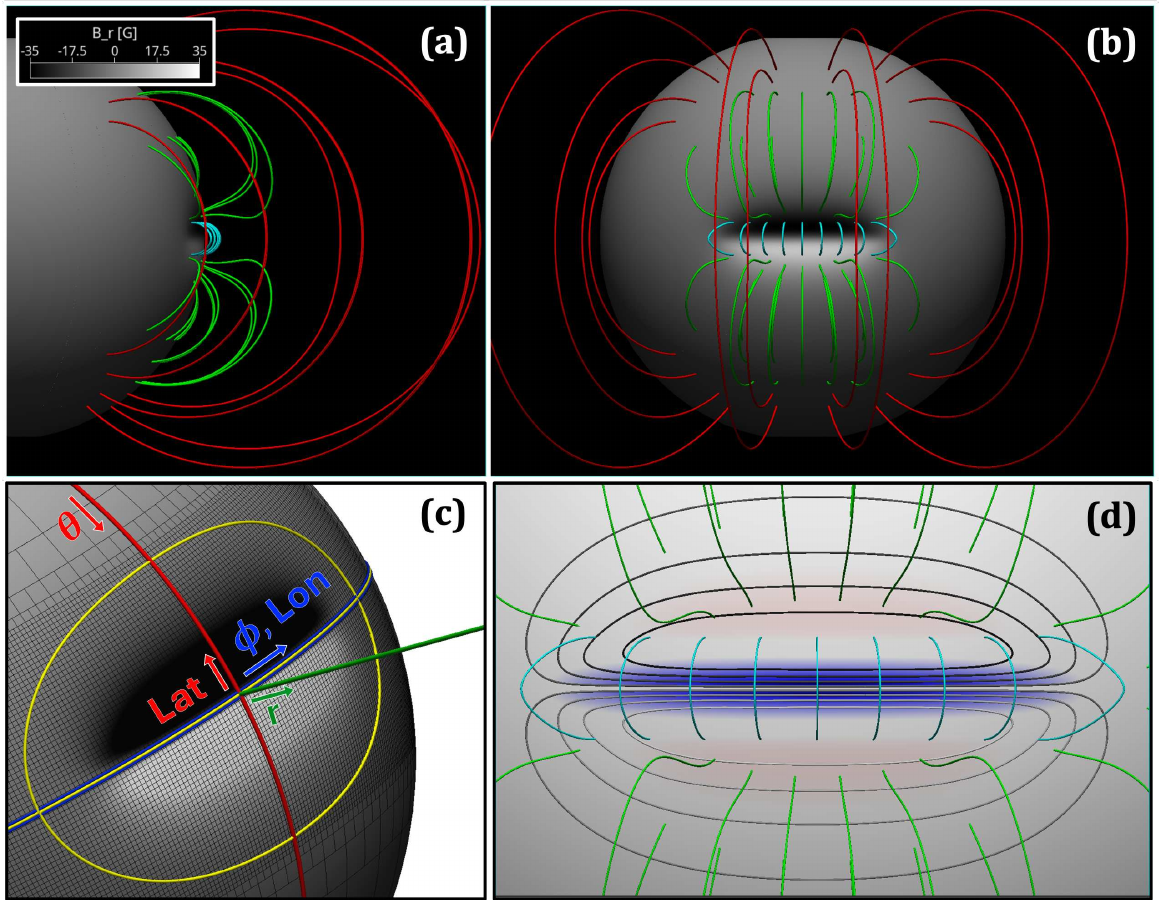}
\caption{Initial configuration and driver for the eruptive flare model. (a-b) Magnetic topology shown from two orthogonal perspectives: (a) east and (b) above. White (black) shading at the inner boundary corresponds to positive (negative) polarity. Four distinct flux regions are indicated in cyan, green, and red (northern and southern sets of green field lines represent distinct flux regions). (c) Illustration of initial solar surface grid (shown at half resolution for clarity) and coordinate system. Green, red, and blue curves indicate the radial, polar $(\theta)$/latitude and azimuth ($\phi$)/longitude coordinates, respectively, with arrows indicating their directions of increase. The polarity inversion lines (PILs) are indicated in yellow, with the oval PIL encircling the active region extending approximately $\pm34^\circ$ in longitude and $\pm22^\circ$ in latitude. (d) Profile of $B_\phi$ component of the tangential flux injected by the STITCH method (see text for details) at the inner boundary. Contours of the radial component of the magnetic field are overlaid in grayscale.
\label{fig:config}}
\end{figure}

The numerical calculations were performed with the Adaptively Refined Magnetohydrodynamics Solver \citep[ARMS;][]{devore08a} that has been employed extensively for CME/eruptive flare modeling \citep{lynch08a,lynch09a,karpen12a,masson13a,lynch16a,dahlin19a,dahlin22a,dahlin22b}. The magnetic configuration is illustrated in Fig.~\ref{fig:config} showing three views of the multipolar `breakout' configuration employed. The magnetic flux distribution is antisymmetric across the equator, so that the central Polarity Inversion Line (PIL) lies along the equator. The corresponding flare current sheet, therefore, will lie above the PIL and be approximately planar at its center. This choice simplifies identification and analysis of plasmoids. The symmetry and simplicity of the magnetic profile results in flare ribbons that (a) are oriented primarily in the east-west direction and (b) exhibit structure due exclusively to the reconnection dynamics and virtually independent of the surface flux distribution.
\par
The initial atmosphere was a spherically symmetric hydrostatic equilibrium with a temperature profile scaling as $T = T_s/r$, a base temperature $T_s = 2 \times 10^6$ K, and pressure
$P_s = 4 \times 10^{-1}$ dyn cm$^{-2}$. 
The ideal MHD equations were solved with an adiabatic
temperature equation. The domain consisted of the spherical wedge, $r \in [1R_s, 30R_s]$, $\theta \in [\pi/16, 15\pi/16]$,
and $\phi \in [-\pi,+\pi]$, where r is the radial coordinate, $\theta$ the polar angle (or colatitude), and $\phi$ the azimuthal angle, and $R_s$ is the solar radius. For many of the figures in this article, we use degrees latitude $(\chi = \pi/2 - \theta) $ and longitude ($\phi$). For these coordinates, the domain extents are $\chi \in [-78.75^\circ, 78.75^\circ]$ and $\phi_{\mathrm{deg}} \in [-180^\circ, 180^\circ]$.
Reconnection was facilitated by grid-scale numerical dissipation that broke the frozen-in flux constraint. We employed 6 levels of adaptive refinement corresponding to a minimum grid cell size of $\Delta r = 1.33$~Mm and 
$R_s \Delta \theta = R_s \Delta \phi = 1.08$~Mm at the inner boundary. The coordinate system and the initial grid are illustrated in Fig.~\ref{fig:config}c.
 
As in \textcite{dahlin22a}, the initially potential magnetic configuration was energized via the injection of tangential flux using the method of STatistical InjecTion of Condensed Helicity \citep[STITCH;][]{mackay14a,mackay18a,lynch21a,dahlin22b}. The spatial profile of the $B_\phi$ injection is shown in Fig.~\ref{fig:config}d. The injection of tangential flux was halted at $t = 11,500$~s, well prior to the onset of the eruptive flare. The magnetic field is line-tied, with all velocity components fixed at zero, across the entire inner radial boundary. Otherwise, the boundary conditions are identical to those employed in previous 3D ARMS eruptive-flare calculations \citep[e.g.,][]{dahlin19a,dahlin22a}.
\par
\begin{figure}[ht!]
\plotone{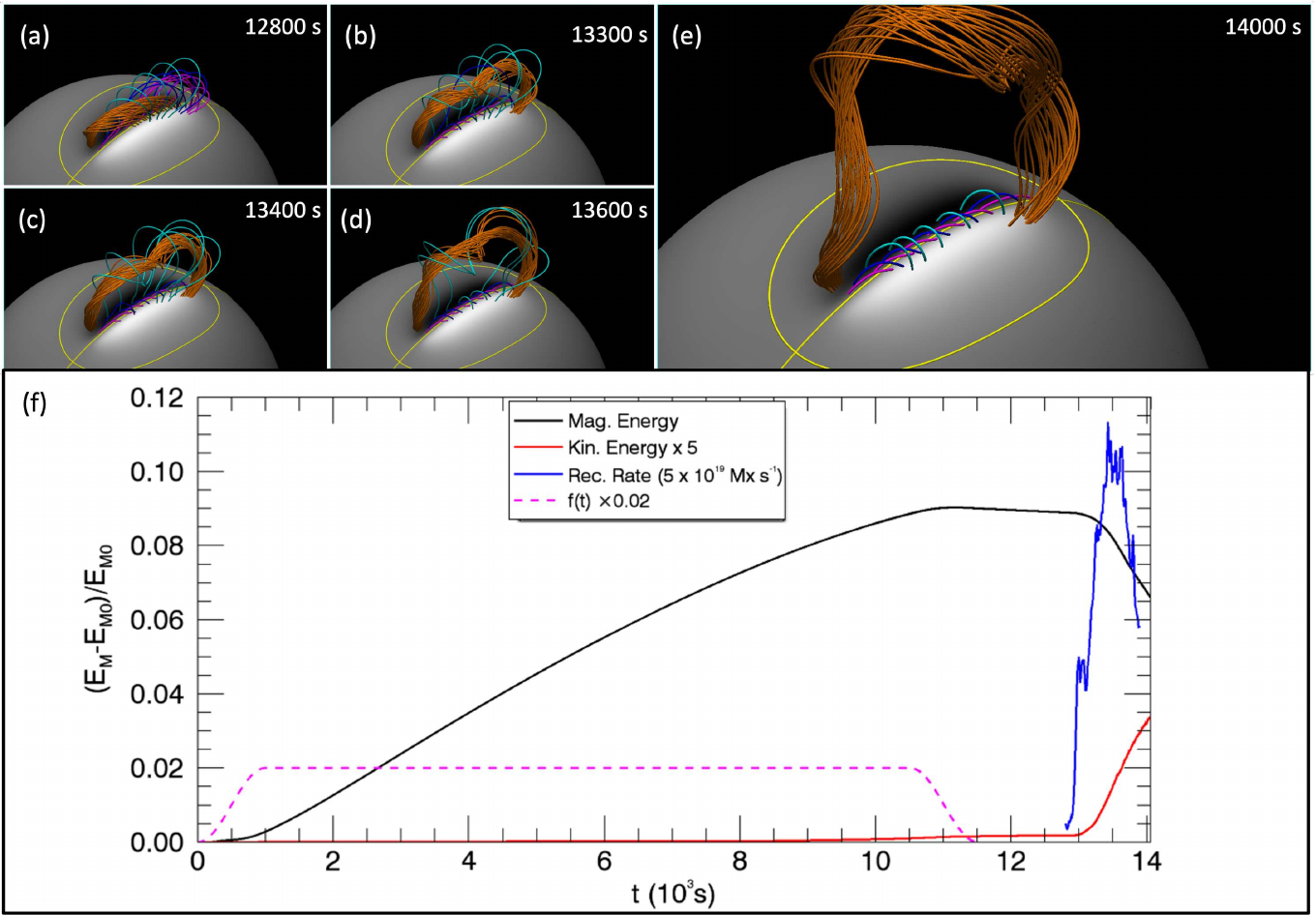}
\caption{Overview of the eruptive flare model evolution. (a-e) Erupting flux rope (orange field lines), overlying flux, and flare loops at five different times during the simulation. Magenta, blue, and cyan indicate field lines progressively further from the PIL (yellow). Grayscale contours show the magnitude of the radial component of the magnetic field. (f) Time-evolution of globally integrated quantities, including the free magnetic energy (black), kinetic energy (red), and reconnection rate (blue). Energies are normalized to the initial potential field energy. The temporal profile for the shear injection is shown in the dashed magenta line. An animation of this figure is available, showing the eruption of the flux rope and formation of the flare arcade during the interval 12,800 s $\leq$ t $\leq$ 14,000 s at a 10 s cadence (the animation duration is 5 s).
\label{fig:eruption_summary}}
\end{figure}

The temporal evolution  of the eruptive flare is summarized in Fig.~\ref{fig:eruption_summary}, which shows five stages of the flare evolution (Fig.\ref{fig:eruption_summary}a-e) and the time evolution of the global energetics (\ref{fig:eruption_summary}f). The orange field lines initially form a sheared arcade (Fig.~\ref{fig:eruption_summary}a) that is subsequently transformed by flare reconnection into a twisted flux rope that erupts outward from the solar surface (Fig.~\ref{fig:eruption_summary}b-e). Simultaneously, the magenta, blue, and cyan field lines that initially overlay the flux rope (Fig.~\ref{fig:eruption_summary}a) stretch outward as the flux rope erupts. Each set of field lines successively reconnects, beginning with the magenta field lines closest to the PIL and proceeding to the blue then cyan field lines (Fig.~\ref{fig:eruption_summary}b-e). A clear strong-to-weak shear transition is evident: the magenta flare loops are highly sheared along the PIL, whereas the cyan field lines are only weakly sheared. This was previously discussed in our investigation of the shear evolution and its link to the reconnection guide field \citep{dahlin22a}, and similar behavior has been noted in many observational flare studies \citep[e.g.][]{su06a,aulanier12a,qiu23a}.

The temporal profile of the STITCH driver and the globally integrated free magnetic energy, kinetic energy, and rate of reconnected flux are illustrated in Figure~\ref{fig:eruption_summary}f. The reconnection rate is calculated via integration of the flux swept out by the flare ribbons (the same method employed in \citet{dahlin22a}). We detail our method for identifying the flare reconnection and the associated ribbons in \S\ref{sec:ribbons}. We emphasize two points illustrated by the global evolution. First, Figure~\ref{fig:eruption_summary}f illustrates that we have captured the self-consistent energy buildup and onset: we began with a potential magnetic field, imposed magnetic shear, and halted the driving at $t = 11,500$~s, well before the onset of fast reconnection ($t \approx 12,800$~s). This ensures that the driving and eruptive phases are well-separated temporally. Second, we note the considerable fluctuations in the reconnection rate (blue) that show the dynamic complexity of the flare reconnection and hint at the presence of plasmoids.

\section{Current Sheet Structure} \label{sec:plasmoids}

\begin{figure}[ht!]
\plotone{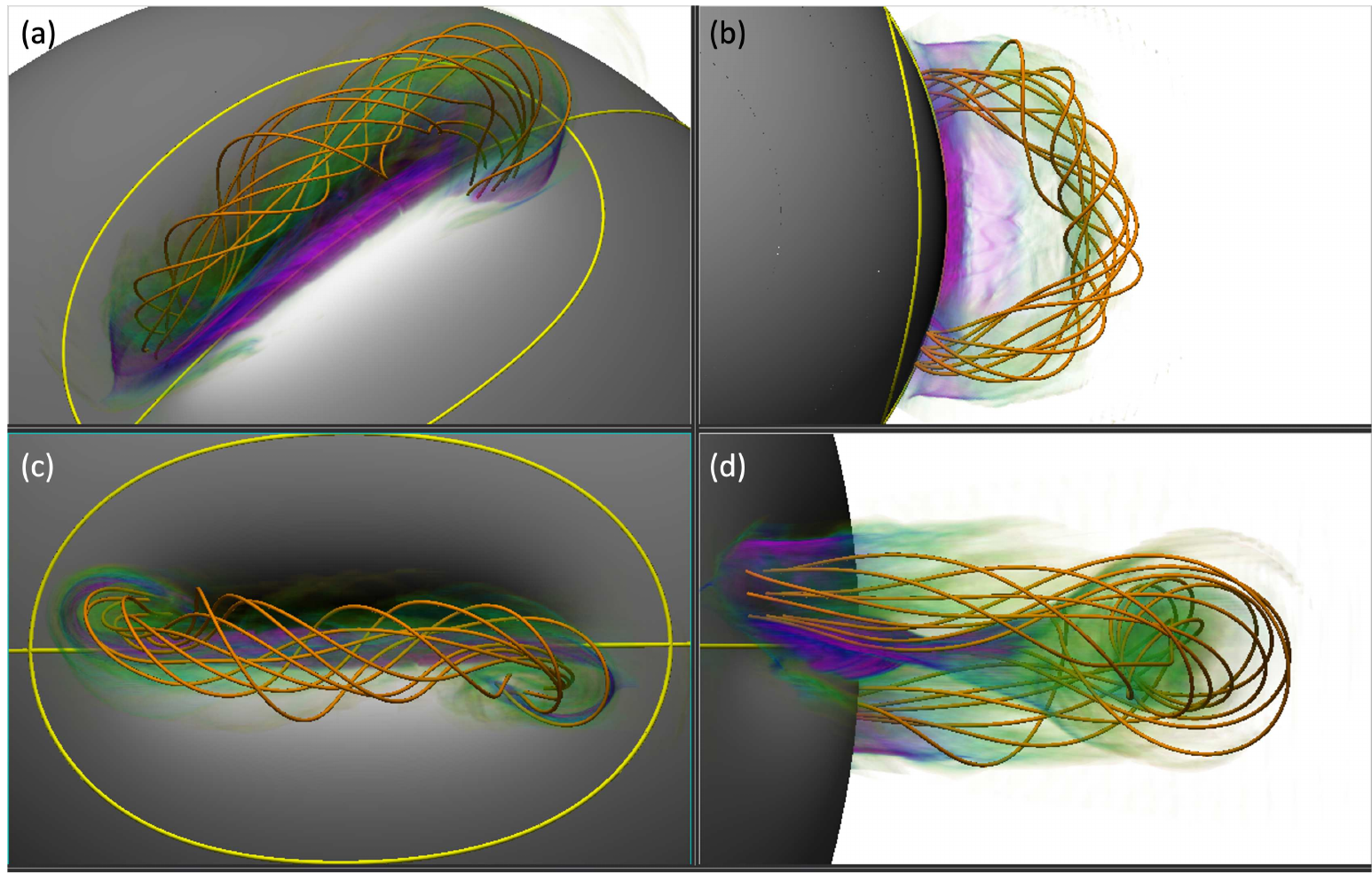}
\caption{Three-dimensional volumetric visualization of the total current density in the flaring region at $t=13,400s$. Magenta indicates the most intense currents. Grayscale color table indicates the radial component of the magnetic field at the inner boundary of the domain, and yellow curves indicate polarity inversion lines. Orange field lines trace the erupting magnetic flux rope.}
\label{fig:flare_context}
\end{figure}

We next examine the fine structure of the flare current sheet. Figure ~\ref{fig:flare_context} shows multiple viewpoints of a volume rendering of the current density at $t = 13,400 s$, when the eruptive flare is well under way. Several notable features are evident in this figure: the most intense current (magenta) has a sigmoidal morphology (Fig.~\ref{fig:flare_context}c), as has been discussed in three-dimensional adaptations of the standard model \citep{janvier14a}. It is also clear that the flux rope winds through a current structure with an oval cross-section (e.g., Fig.~\ref{fig:flare_context}d).
Finally, the current density is highly structured beneath the flux rope, most evident from the viewpoint face-on to the current sheet (Fig.~\ref{fig:flare_context}b), which suggests the presence of plasmoids.
\par
\begin{figure}[ht!]
\plotone{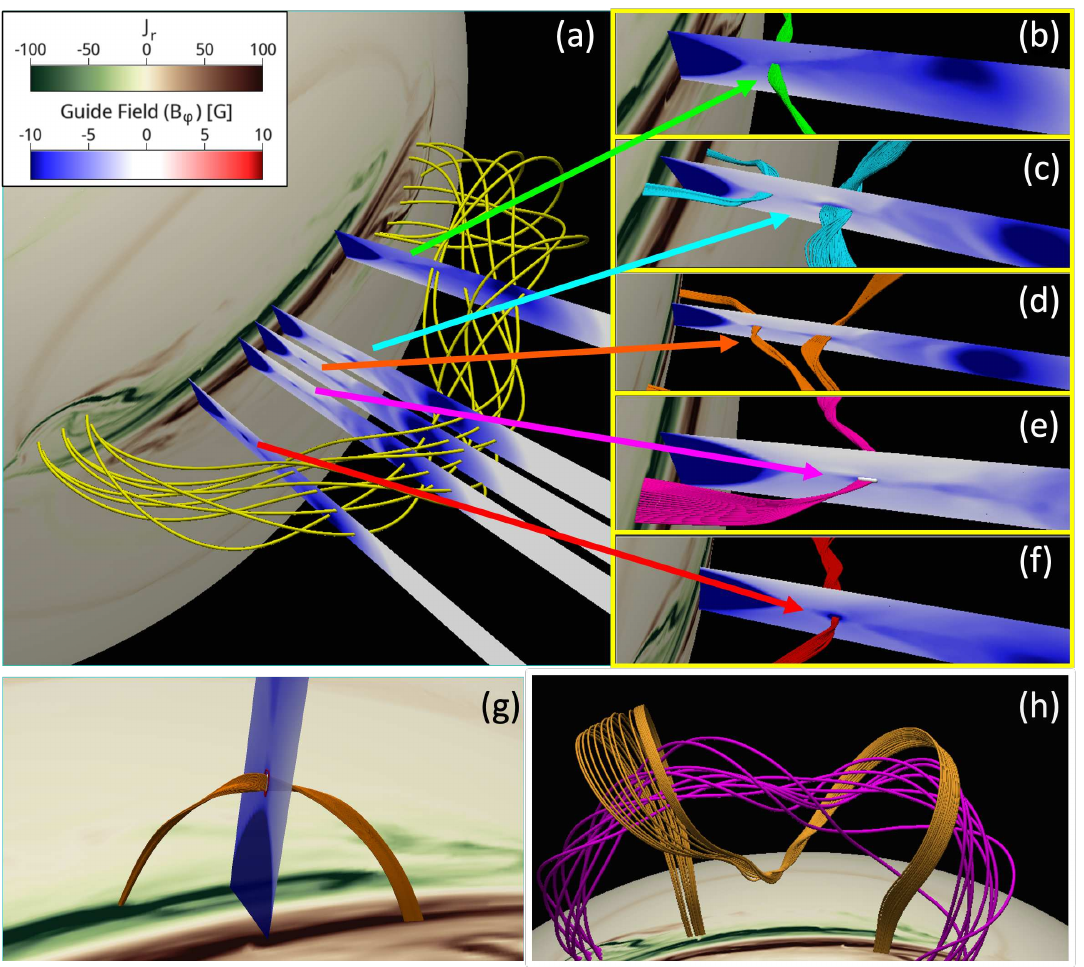}
\caption{Three-dimensional plasmoid structures in the flare current sheet. (a) Context image showing contours of the vertical current density at the inner boundary, illustrating the flare ribbon locations. Yellow field lines indicate the location of the erupting CME flux rope. Five 2D slices of the flare current sheet are also shown, with a color table indicating the guide magnetic field component ($B_\phi$); concentrations of the guide field indicate plasmoids. Panels (b-f) show distinct sets of field lines traced from the slices indicated in panel (a), illustrating the twist and topology of the plasmoids. Examples of plasmoids sunward and anti-sunward of the flare reconnection X-line are shown in (g) and (h), respectively.}
\label{fig:plasmoid_types}
\end{figure}

The detailed magnetic structure is illustrated Fig.~\ref{fig:plasmoid_types}a, which shows the vertical current density at the solar surface, flux rope field lines, and several slices of constant $\phi$ with plots of the strength of the guide field $(B_\phi)$. A significant guide field is expected during the main phase of a flare \citep{qiu17a,dahlin22a,qiu23a}, and upon advection into the current sheet, compression inside plasmoids generates oval-shaped enhancements of the guide field. This makes $B_\phi$ an effective proxy for identifying cross-sections of plasmoid cores. Multiple such concentrations are evident in the five constant-$\phi$ slices in Fig.~\ref{fig:flare_context}a. 
\par
To further illustrate that the $B_\phi$ concentrations are, indeed, plasmoids (i.e., exhibit coherent small-scale magnetic twist), we trace several sets of field lines from the regions of $B_\phi$ enhancement (Fig.~\ref{fig:plasmoid_types}b-f). As expected, each bundle of field lines so identified exhibits significant twist near the plane from which they were traced. It is well established that the  formation and interaction of many such structures throughout the current layer may result in a highly complex, stochastic magnetic topology \citep[e.g.,][]{daughton11a,liu13a_y,dahlin15a,dahlin17a}. However, in order to draw the clearest possible link between plasmoid structures and ribbon features, we focus on two limiting topological cases for the plasmoids. Subsequent to transient initial dynamics near the primary reconnection X-line, the plasmoids propagate `downstream' toward either the sunward (Figure ~\ref{fig:plasmoid_types}g) or anti-sunward direction (Fig.~\ref{fig:plasmoid_types}h). The structure of the former manifests as a twisted segment along a flare loop, whereas the latter behavior corresponds to a short twisted segment along a longer flux tube that wraps around the CME flux rope. These represent the simplest 3-D generalizations of simplified 2.5-D flare plasmoids, which would be detached from both the CME flux rope and the solar surface. Critically, however, these plasmoids are magnetically connected to the solar surface at both ends. We will show that this connectivity imprints complex fine structure in the reconnection topology and, hence, along the flare ribbons.

\section{Model Analogues of Flare Ribbons} \label{sec:ribbons}

As is commonly found in previous ARMS modeling \citep{lynch13a,edmondson17a,dahlin22a}, the highly-resolved flare current sheet of the present simulation is characterized by a proliferation of plasmoids, which are of such spatial and temporal scales that direct measurement would be highly challenging. Furthermore, fully rigorous forward modeling of flare ribbons would require accurate modeling of the flare energy release, transport, and loss processes. Given the observation of intense particle acceleration, these processes are likely to involve kinetic effects. Consequently, such forward modeling is well-beyond present theoretical and computational capabilities. We therefore instead endeavor to identify a useful qualitative proxy for the flare ribbons, prioritizing the identification of fine-scale structure due to magnetic field dynamics in the current sheet. 
\par
Given that flare ribbons are understood to be the locations where energy released by reconnection is deposited impulsively (e.g. by particle beams), we may proceed with the simplifying approximation that ribbons are the instantaneous footpoints of flare-reconnected field lines. However, identifying individual reconnection events in 3-D numerical studies can be highly challenging, given that flux surfaces are easily broken by turbulent dynamics. Furthermore, whereas in 2-D geometries reconnection involves discontinuous topology changes across separatrix layers, in three-dimensional systems a generalized form of reconnection involving continuous changes in connectivity may occur \citep{schindler88a}, often referred to in flare contexts as `slipping' reconnection \citep[e.g.,][]{priest95a}.
\par
Despite these challenges, substantial progress in diagnosing 3-D reconnection has been achieved through analysis that aims to quantify the changes in magnetic topology. For example, the squashing factor \textit{Q}, \citep[e.g.,][]{demoulin96a,titov02a} characterizes deformations in field-line mapping, which may be used to identify `quasi-separatrix layers' where the gradient of the magnetic connectivity is strong and reconnection may preferentially occur. Another useful quantity correlated with the location of flare ribbons is the vertical current density. As described in detail in \textcite{janvier14a}, the flare current sheet maps to vertical current concentrations at the inner boundary, so that as the reconnection proceeds and the flare current sheet rises, the flare ribbons separate in time. A useful aspect of the vertical current as a proxy for ribbons is that it may be directly calculated in both observations and in simulations.
\par
\begin{figure}[ht!]
\plotone{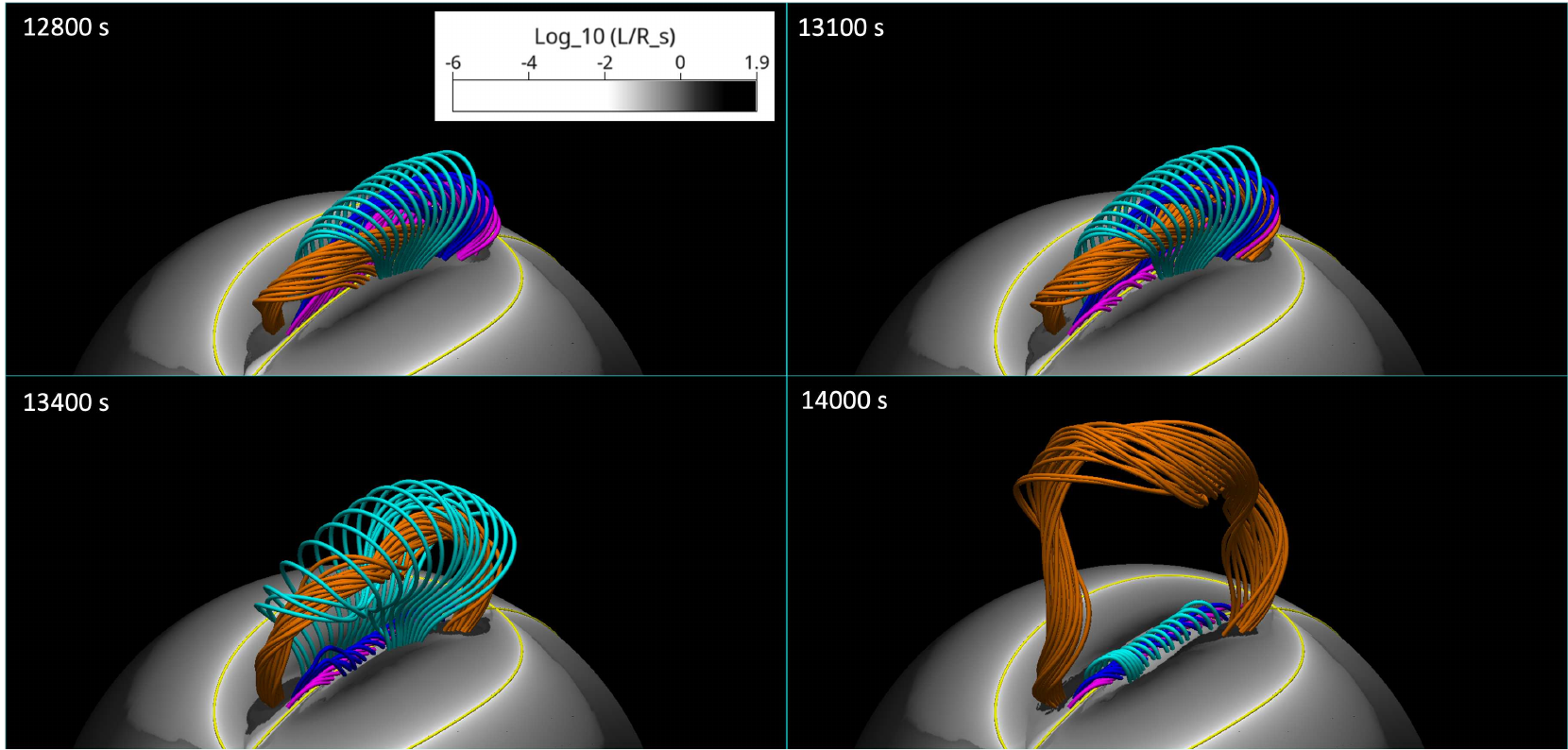}
\caption{
Illustration of the link between flare reconnection and field line length length at four different times during the evolution of the eruptive flare (the times are identical to those shown in Fig.~\ref{fig:eruption_summary}). The field-line length map (\textit{L}-map) is shown at the inner boundary ($r=R_s$) where white/light gray indicate short field lines and black/dark gray indicate long field lines. Yellow curves indicate PILs. The magenta, blue, and cyan field lines are identical to those shown in Fig.~\ref{fig:eruption_summary}. As flare reconnection proceeds, these field lines expand outward, then reconnect, and form the flare arcade. The resulting lengths of the associated field lines abruptly shorten, as reflected in the white region that expands outward from the central PIL. Simultaneously, the footpoints of the orange field lines (representing the erupting twisted flux rope) darken, corresponding to the outward expansion of the flux rope. An animation of this figure is available, showing the evolution of the $L$-map on the solar surface during the eruption of the flux rope and formation of the flare arcade during the interval 12,800 s $\leq$ t $\leq$ 14,000 s at a 10s cadence (the animation duration is 5 s).
}
\label{fig:lmap_3d}
\end{figure}

Although useful, these measures are not direct signatures of field-line connectivity changes due to reconnection, and carry significant limitations for analysis of fine structures as we will illustrate below. Instead, we employ the concept of a `field-line length map', or $L$-map, which is constructed by tracing a field line upward from the solar surface to its other end, then assigning the value of its length to the originating footpoint (see further discussion in the Appendix).
The utility of the $L$-map is illustrated by the series of snapshots shown in Fig.~\ref{fig:lmap_3d}. The region of shortened field lines (white) near the central PIL corresponds to the flare arcade; its boundary represents the most recently reconnected flare ribbons. Indeed, the successive reconnection of the magenta, blue, and finally cyan field lines corresponds closely to the spreading of the region of shortened field lines. 
\par
\begin{figure}[ht!]
\epsscale{0.8}
\plotone{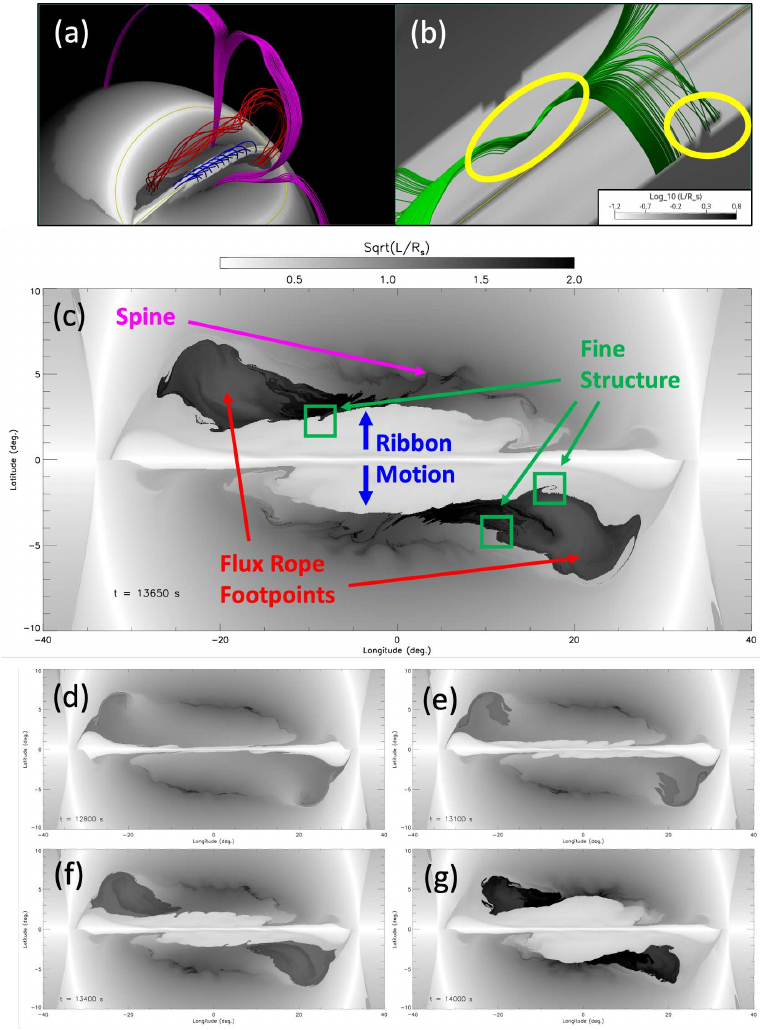}
\caption{
Illustration of important features revealed by an $L$-map. A three-dimensional representation of four distinct sets of field lines (a) is compared to a notated high-resolution $L$-map (c). Red indicates flux rope field lines and corresponding footpoint regions. Blue indicates the flare arcade, the boundary of which is an analogue for flare ribbons. Magenta denotes long field lines that pass near coronal null points. Green field lines in (b) show a twisted flux bundle (plasmoid) that maps to fine structure in the $L$-map. Several such regions are highlighted in the middle panel. Panels (d)-(g) show four snapshots of the $L$-map corresponding to the times shown in Fig.~\ref{fig:lmap_3d}. An animation of the high-resolution $L$-map is available, showing the $L$-map evolution during the flare which involves outward spreading of the flare arcade indicated by shortened field lines, the expansion of the flux rope indicated by lengthening field lines, and the development of fine structure. The animation covers the interval 12,800 s $\leq$ t $\leq$ 14,000 s at a 10 s cadence (the animation duration is 5 s).
\label{fig:lmap_explainer}}
\end{figure}

By contrast, regions of lengthening field lines (dark gray) correspond to the footpoints of the erupting flux rope. The growth of these regions therefore indicates the ongoing addition of magnetic flux via the flare reconnection, and the lengthening of field lines within this region corresponds to the outward expansion of the flux rope. Key features of the $L$-map are shown in Fig.~\ref{fig:lmap_explainer}, which illustrates how the flare arcade, flux rope footpoints, and the footpoints of spine lines may be discerned in an $L$-map. Crucially, fine structure present at the leading edge of the field-line shortening corresponds to the footpoints of a plasmoid in the current sheet (we examine such structure in detail in \S\ref{sec:structure}). These results demonstrate that field-line length maps ($L$-maps) are highly useful for identifying the footprints of reconnected field lines and, hence, deriving a proxy for flare ribbons.

\begin{figure}[ht!]
\plotone{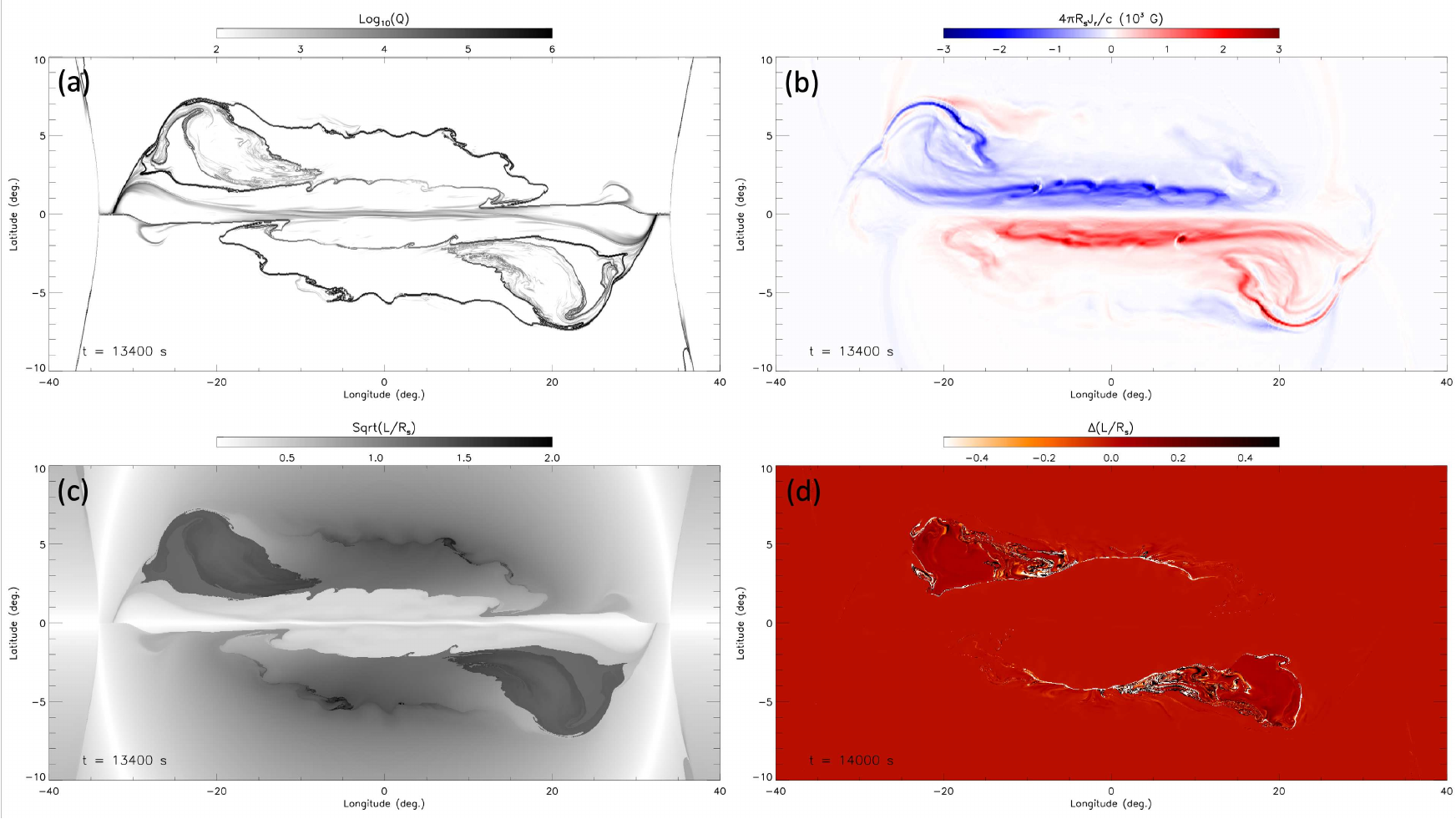}
\caption{Three proxies for flare ribbons. (a) Squashing factor ($Q$), denoting strong gradients in magnetic field connectivity. (b) Vertical current density ($J_r$, marking the footprints of the flare current sheet on the solar surface. (c) Field-line length maps, indicating distinct topological regions associated with discontinuous, contrasting field-line lengths. (d) A simple proxy for flare ribbons given by $\Delta L(t) = L(t + 10) - L(t - 10)$.}
\label{fig:ribbonproxies}
\end{figure}

In Figure~\ref{fig:ribbonproxies}a-c we compare high-resolution images of the squashing factor $Q$, vertical current $J_r$, and the field-line length $L$ ($Q$ and $L$ are calculated using the ARMS suite of topology tools described in \citet{wyper21a}). Most of the features illustrated in Fig.~\ref{fig:lmap_explainer} can be identified using each of these three quantities (the exception is that $J_r$ does not show the base of the spine). For example, the boundary of the flare arcade in Fig.~\ref{fig:ribbonproxies}c (the ribbon front) located near $\pm 2$ deg. latitude corresponds to both a high-$Q$ layer (\ref{fig:ribbonproxies}a) and a region of strong $|J_r|$ (\ref{fig:ribbonproxies}b). Indeed, \textcite{tassev17a} have previously noted that the field-line length can be useful as an efficient `quick and dirty'  method to identify regions likely to host QSLs. Likewise, fine structure can be discerned along the ribbon fronts for all three quantities. Even so, as was illustrated in Fig.~\ref{fig:lmap_explainer}, the clearest delineation of topological regions is given by the $L$-map. 
\par
For the purpose of identifying flare ribbon fine structure, $Q$ and $J_r$ are further limited in utility with respect to the $L$-map. The vertical current is calculated in terms of gradients of $\mathbf{B}$ and therefore subject to significant noise and furthermore constrained by the grid resolution. Indeed, fine-scale structures in Fig.~\ref{fig:ribbonproxies}a,c are more challenging to identify in Fig.~\ref{fig:ribbonproxies}b where they instead show up as intense current blobs. A significant drawback of the squashing factor, in turn, is that while many of the fine structures are captured, it is computationally expensive to calculate, because a dense set of field lines is required in order to calculate accurately the gradients in the mapping. This can be problematic in the case of small plasmoids. Furthermore, since it does involve calculating gradients, it is inherently more sensitive to numerical noise in the results (see for example the flux rope footpoints at $\pm$20 degrees longitude, $\mp$ 5 degrees latitude in Fig.~\ref{fig:ribbonproxies}a).
\par
Further challenges arise when aiming to use either quantity for deriving ribbon proxies. Features in either $J_r$ or $Q$ must be both (a) positively identified as the base of the flare current sheet and (b) resolved with sufficient temporal resolution to capture the moment that a features sweeps across a point of interest. The regions of large Q illustrated in Fig.~\ref{fig:ribbonproxies}a correspond to a variety of categories of topological structures, which are not limited to flare ribbon fronts but also include the base of the spine, the boundary of the flux rope footpoints, and structure within the flux rope itself. Therefore, \textit{Q} on its own is not sufficient to identify locations of flare reconnection. Furthermore, \textit{Q} only indicates strong connectivity gradients where reconnection might be \textit{likely} to occur, but does not itself identify reconnection sites. With regard to the vertical current density, while it is an intuitive quantity for mapping of the flare current sheet to the lower solar atmosphere, similar to \textit{Q} it neither disambiguates between different types of current sheets nor definitively indicates reconnection sites, and would also require a cadence sufficient to resolve the 'moment' of reconnection at a given location.
\par
By contrast, \textit{L} is \textit{specific} in that it can be used to classify the relationships of the topological regions (i.e., $L_{\texttt{fluxrope}} > L_{\texttt{overlying}} >L_\texttt{arcade}$) and therefore the type of reconnection associated with the motion of the boundaries between these regions. Additionally, $L$ is \textit{robust} in that it is comparatively insensitive to the time cadence (the reconnection event itself does not need to be resolved, as it is only necessary to show that $L$ has changed within a given interval) as well as the spatial resolution of the map (no gradients are computed). Furthermore, reconnection associated with a given location requires minimal additional context beyond the local $L(t)$ (for example, the typical field-line length in each region, for use in selecting appropriate thresholds for identification of reconnection events).
\par
\begin{figure}[ht!]
\plotone{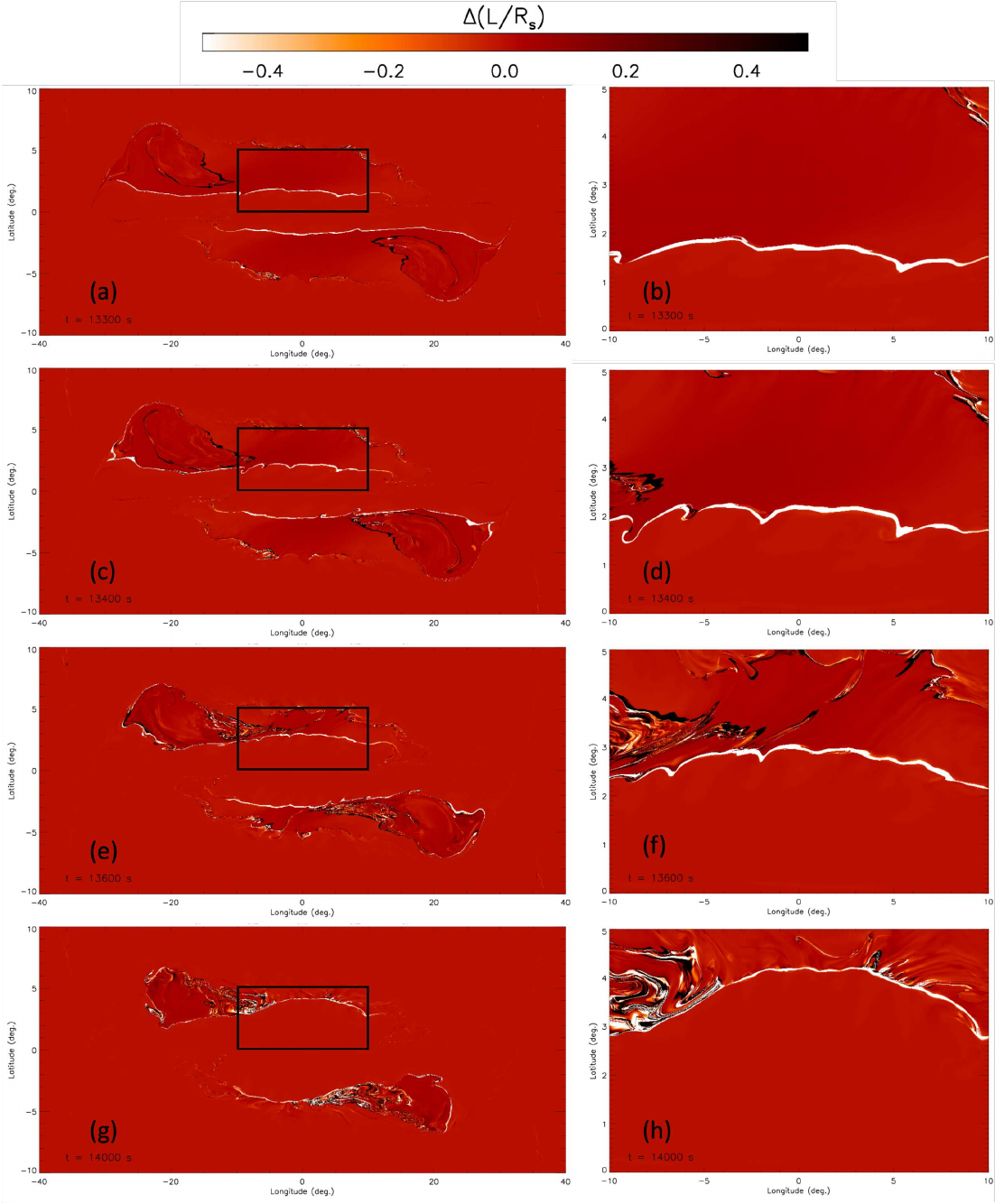}
\caption{Rudimentary proxy for flare ribbon fronts, illustrated via running difference images of $L$ at different times throughout the flare evolution, calculated for a given time as $\Delta L(t) = L(t + 10) - L(t - 10)$. The left panels depict the entire flare region, whereas the right panels show zoomed-in regions corresponding to the rectangles indicated in the context figures. Light colors (white-yellow) indicate field-line shortening corresponding to flare reconnection (a proxy for the location of the flare ribbon front), whereas dark colors (red-black) indicate field-line lengthening. The left panels illustrate the typical large-scale morphology of flare ribbons including the outward spread and development of ribbon hooks. The right-hand panels show many fine-scale features (e.g., spirals, breaking waves) along the ribbon front. Separate animations of the $L$-map running difference sequence are available for both the large-scale and inset regions indicated in this figure over the interval 12,800 s $\leq$ t $\leq$ 14,000 s at a 10s cadence. The context animation shows the flare-ribbon-like spread of the field-line shortening, and the inset region highlights the spiral-like structure along a patch of the ribbon front (the duration of both animations is 5 s).}
\label{fig:ribbon_proxy_4frames}
\end{figure}

The utility of the $L$-map is further illuminated in a running difference image (Fig.~\ref{fig:ribbonproxies}d). Field lines that have shortened in the previous time interval ($\Delta L < 0$) correspond to the footpoints of reconnected field lines, colorized in yellow/white to evoke the interpretation that these are locations where flare ribbons would form (we present a more sophisticated method for constructing synthetic ribbons in \S\ref{sec:observations}).
Figure ~\ref{fig:ribbon_proxy_4frames} shows $L$-map running difference images at four different times. The large-scale evolution (left panels) reflects an expansion outward of the ``ribbons'' from the central PIL. The right-hand panels, which zoom in on a smaller region of the ribbons, highlight numerous fine-scale structures that form along the ribbon frontier. These structures primarily take the form of spirals/breaking waves (as was previously discussed in the analytical model presented in WP21). 
\par
For the reasons presented in this section, in the following analysis we focus exclusively on the properties of the $L$-maps in order to investigate the relationship between the flare ribbon structure and current sheet dynamics. Before doing so, we we wish to briefly note the broader potential for interpreting flare dynamics in terms of the evolution of the field-line length. In addition to `brightenings', the running difference images also reveal locations likely to be associated with coronal dimming ($dL/dt > 0$, indicated by dark colors). The lengthening of field lines in the flux rope footpoints correspond well to the well-known coronal dimmings that are thought to result from expansion or opening up of the magnetic field \citep[e.g.,][]{sterling97a,qiu24a}, and lengthening of field lines just outside the flare arcade may be associated with subtler dimmings resulting from the stretching of field lines prior to flare reconnection, as was recently pointed out in \textcite{qiu24a}. Other examples of reconnection can also be identified via $L$-maps and/or running difference images, which include the case of filament threads reconnecting to form flare loops \citep{lorincik19a}, reconnection of the flux rope with itself \citep[filament leg-leg reconnection,][]{dudik22a}, as well as reconnection associated with the base of the spine/breakout current sheet (which may involve any of the three highlighted regions: flux rope, overlying field, and flare arcade).
\par

\par

\section{Ribbon Fine Structure} \label{sec:structure}

\begin{figure}[ht!]
\plotone{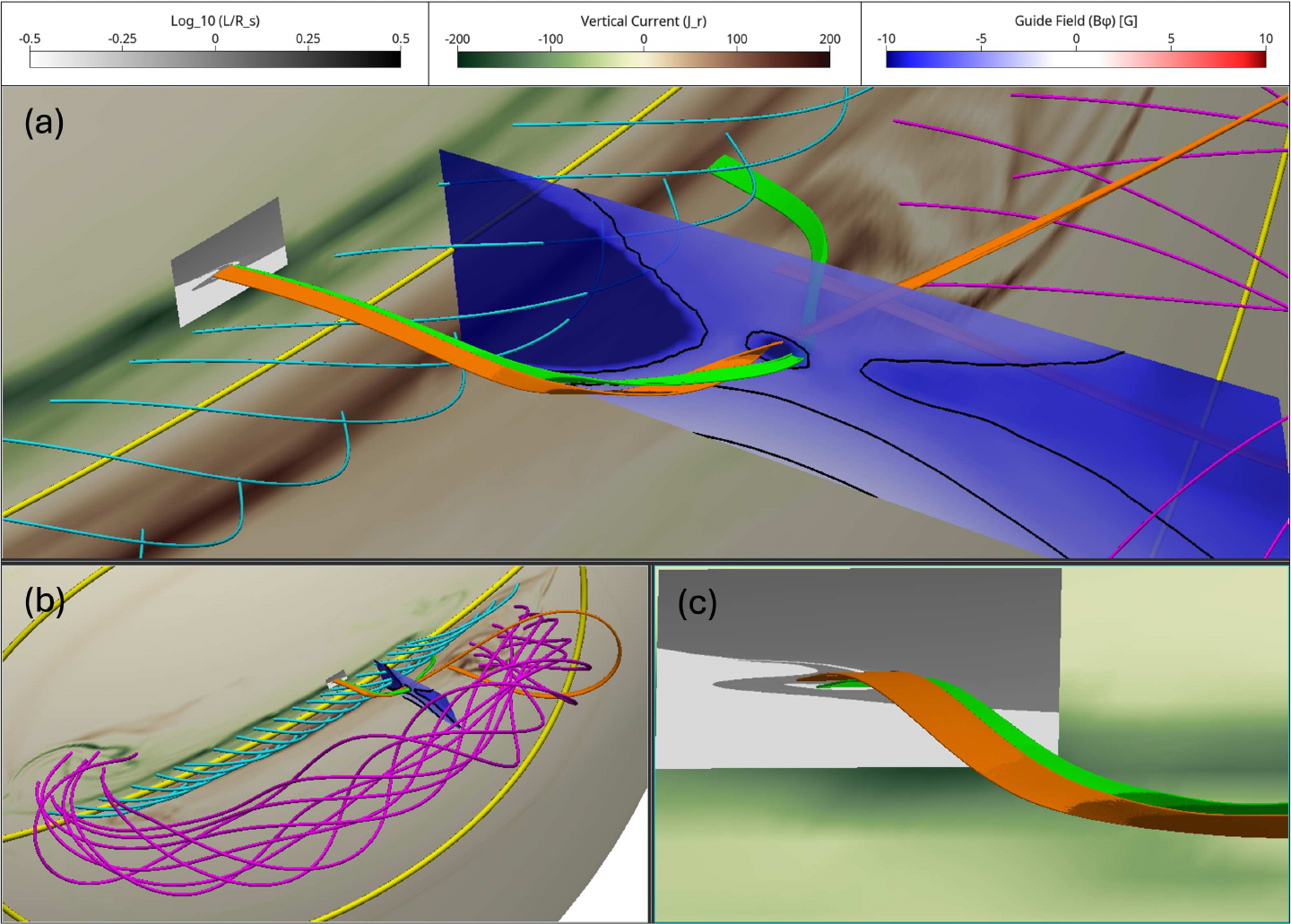}
\caption{Illustration of the physical connection between a plasmoid and its topological imprint on the solar surface at $t = 13340$~s.
Panel (a) shows two field line strips traced from a small-scale spiral on the solar surface $L$-map (grayscale) to a flare current sheet plasmoid (blue oval in a longitude cut $\phi = 8.5^\circ$ of the guide field). The vertical current at the solar surface ($J_r$, green-brown colormap) as well as the flare arcade (light blue field lines), the CME flux rope (magenta field lines), and the PILs (yellow) are shown for context. The green field line strip was traced from the light gray region of the spiral (shorter field lines) and maps to a conjugate region in the positive polarity region, whereas the orange field line strip was traced from the dark gray region (longer field lines) and wraps once about the CME flux rope, mapping to a region along the flux rope hook in the positive polarity region. Panel (b) shows a zoomed-out view of (a) illustrating the location of the plasmoid within the eruptive flare/CME configuration. Panel (c) shows a zoomed-in view of the solar surface $L$-map patch and the footpoints of the orange and green field-line strips. The temporal evolution of the guide field cut and the $L$-map patch is shown in Fig.~\ref{fig:merger_down_6frames}.
\label{fig:plasmoid_spiral}
}
\end{figure}

We have shown that (a) the model current sheet is replete with many plasmoids and (b) our model analogue of flare ribbons is likewise characterized by copious fine structure. We now proceed to investigate the nature of the physical link between these two features. We first present a snapshot of a plasmoid and its solar surface imprint in Fig.~\ref{fig:plasmoid_spiral} in order to illustrate how the localized magnetic twist of the plasmoid distorts the topology of the flare ribbon. We then proceed to explore the temporal evolution of two representative plasmoids to illustrate how the dynamics impact the $L$-map signatures.

A snapshot of a plasmoid and its associated $L$-map structure is shown in Fig.~\ref{fig:plasmoid_spiral}. Regions of long (dark gray) and short (light gray) field lines wrap about each other in a `spiral' pattern (similar to what was shown in WP21). Field line strips traced from the interior arcs of the spiral twist about each other, passing through a plasmoid (oval feature in the guide-field slice) before separating as the green strip maps directly to the solar surface, whereas the orange field strip (shown in context in Fig.~\ref{fig:plasmoid_spiral}b) wraps around the erupting flux rope and maps to a section of the ribbon hooks (visible in $J_r$). The $L$-map feature thereforeillustrates how twisted field-line bundles in the corona distort the boundary between magnetic field lines that have already (green) or not yet (orange) fully reconnected to become a part of the flare arcade. We note that the spiral is reminiscent of features that might result from the action of a Kelvin-Helmholtz instability (KHI). However, whereas the shear flows involved in the KHI roll up an interface between two fluids and is reflected in the local plasma properties, the $L$-map instead encodes non-local topological information. It is clear from Fig.~\ref{fig:plasmoid_spiral}a,c that the field line strips do not begin to twist and wrap about each other until well into the corona. Indeed, the line-tying boundary conditions imply that the spiral can only formed via \textit{topological changes} along the field lines (i.e., reconnection associated with the plasmoid formation).

\begin{figure}[ht!]
\plotone{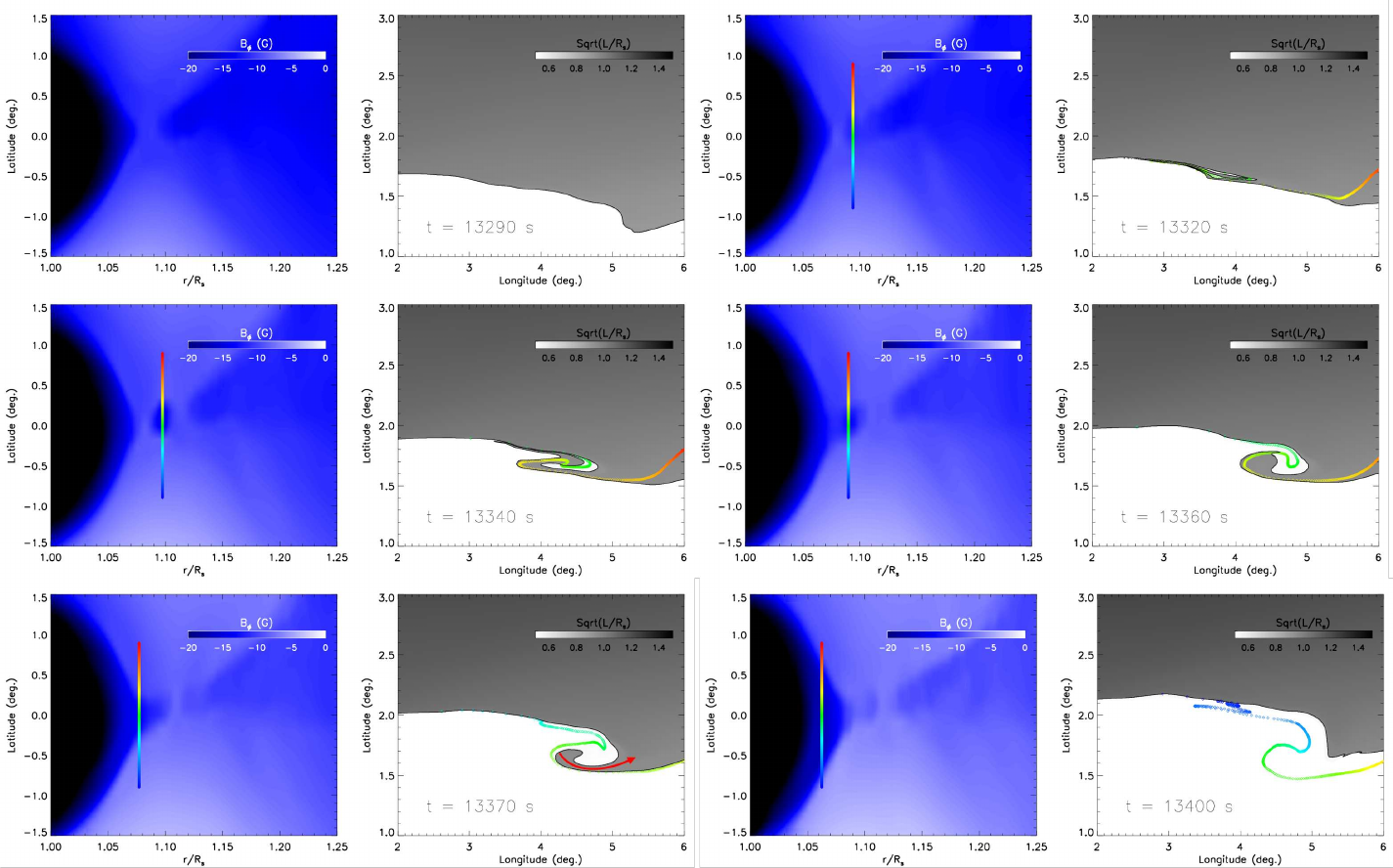}
\caption{
Evolution of a downward-propagating plasmoid and its mapping to the solar surface. For each time step shown, the left-hand panels show the guide field $B_\phi$ on a 2D slice through the flare current sheet corresponding to $\phi = 8.5^\circ$. The rainbow-colored slit indicates the seed points for a set of field lines traced to the solar surface. The right-hand panels show a localized patch of the $L$-map at the solar surface, with the light region corresponding to the footprint of the flare arcade and the dark region corresponding to overlying magnetic flux that has not yet reconnected. The boundary between these two regions is a proxy for the ribbon front. Rainbow symbols show where the field lines traced from the plasmoid map to the $L$-map. An animation of the mapping from the plasmoid cross-section in the corona to the solar surface is available, covering the interval 13,290 s $\leq$ t $\leq$ 13,400 s at a 1 s cadence. The animation shows the formation, propagation, and merger of a plasmoid into the looptop and the concurrent formation and `filling-in' of a spiral feature in an $L$-map patch. (the animation duration is 5 s).
\label{fig:merger_down_6frames}
}
\end{figure}

\begin{figure}[ht!]
\plotone{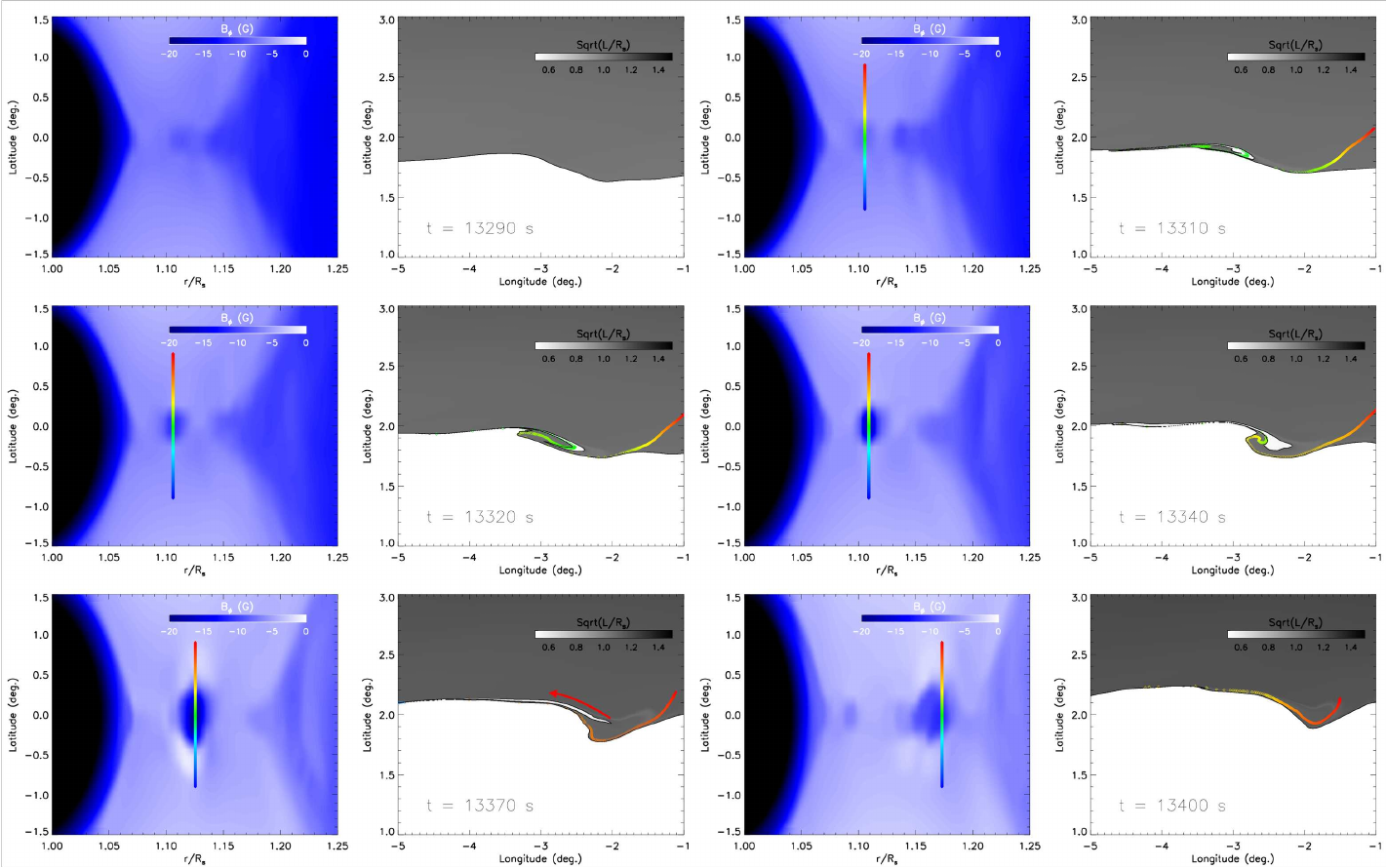}
\caption{
Evolution of an upward-propagating plasmoid and its mapping to the solar surface. For each time step shown, the left-hand panels show the guide field $B_\phi$ on a 2D slice through the flare current sheet corresponding to $\phi = 0.4^\circ$. The rainbow-colored slit indicates the seed points for a set of field lines traced to the solar surface. The right-hand panels show a localized patch of the $L$-map at the solar surface, with the light region corresponding to the footprint of the flare arcade and the dark region corresponding to overlying magnetic flux that has not yet reconnected. The boundary between these two regions is a proxy for the ribbon frontier. Rainbow symbols show where the field lines traced from the plasmoid map to the $L$-map.  An animation of the mapping from the plasmoid cross-section in the corona to the solar surface is available, covering the interval 13,290 s $\leq$ t $\leq$ 13,400 s at a 1 s cadence. The animation shows the formation and upward propagation of a plasmoid toward the CME flux rope and the concurrent formation and retraction/unwinding of a spiral feature in an $L$-map patch. (the animation duration is 5 s).
\label{fig:merger_up_6frames}}
\end{figure}

Having established the premise of the physical connection between plasmoids and $L$-map fine structure, we now examine the dynamics associated with the plasmoid evolution. Figures~\ref{fig:merger_down_6frames} and \ref{fig:merger_up_6frames} show the evolution of two representative plasmoids (corresponding to Fig.~\ref{fig:plasmoid_types}g,h respectively) and their corresponding mappings to the solar surface. In both cases, a slice of $B_\phi$ through the current sheet is shown in order to indicate the locations of plasmoids, the cores of which correspond to oval concentrations of $B_\phi$ (as previously shown in Fig.~\ref{fig:plasmoid_types}). For the downward-propagating plasmoid, these panels correspond exactly to the $L$-map and guide field cuts shown in Fig.~\ref{fig:plasmoid_spiral}a. A selection of field lines  are traced from the current sheet (rainbow slit) to the solar surface (rainbow symbols). The $L$-map patch linked to the core of the plasmoid is shown for each time step.
\par
We focus first on Fig.~\ref{fig:merger_down_6frames}, an example of a downward-propagating plasmoid that forms near $\phi \approx 8.5^\circ$ (as indicated in Fig.~\ref{fig:plasmoid_types}g) and Fig.~\ref{fig:plasmoid_spiral}. In the first frame ($t = 13,290s$), no clear plasmoid is visible and the $L$-map patch shows a smooth boundary between long and short field lines. The first hints of a plasmoid appear in $t=13,320$ s. Its footpoints are interwoven with a small, flattened spiral that has begun to form on the ribbon front in the $L$-map. By $t= 13,340$ s both the plasmoid and the corresponding spiral have fully formed. The plasmoid then begins to propagate downward ($13,360$ s) while the spiral begins to fill in (indicated by a red arrow in Fig.~\ref{fig:merger_down_6frames}). Finally, the plasmoid merges fully into the looptop ($t = 13,400$ s) and the $L$-map once again has a smooth boundary between the reconnected and unreconnected flux, as was the case at $t= 13,290$ s, although the flare arcade has spread to a higher latitude. Interestingly, the warping of the field line mapping remains evident in the wavelike feature formed by the footpoints shown at $t = 13,400$ s.
\par

\par

We contrast this evolution with that of an upward-propagating plasmoid (Fig.~\ref{fig:merger_up_6frames}, also shown in Fig.~\ref{fig:plasmoid_types}h). This plasmoid is located near the center of the flare configuration at $\phi \approx 0.4^\circ$ (the guide field weakens earlier here \citep{dahlin22a}, which is why the ambient $|B_\phi|$ is smaller than for the example shown in Fig.~\ref{fig:merger_down_6frames}).  A $B_\phi$ enhancement develops and begins to propagate upward, growing significantly larger than its counterpart shown in Fig.~\ref{fig:merger_down_6frames}. This occurs because the propagation time is longer (the plasmoid must catch up to the erupting flux rope) and the magnetic field is weaker (so that the plasmoid expands as it rises). However, while an $L$-map spiral forms (e.g., $t = 13,320$ s), it is significantly smaller than that associated with the downward-propagating plasmoid (Fig.~\ref{fig:merger_down_6frames}), despite the larger cross-section of the upward-propagating plasmoid. Furthermore, instead of filling in with reconnected flux (white) as in Fig.~\ref{fig:merger_down_6frames}, the spiral appears instead to `unwind' as a light-colored arc of the spiral retracts (indicated by the red arrow in Fig.~\ref{fig:merger_up_6frames}) even while the area of the plasmoid cross section continues to grow.
\par
\begin{figure}[ht!]
\plotone{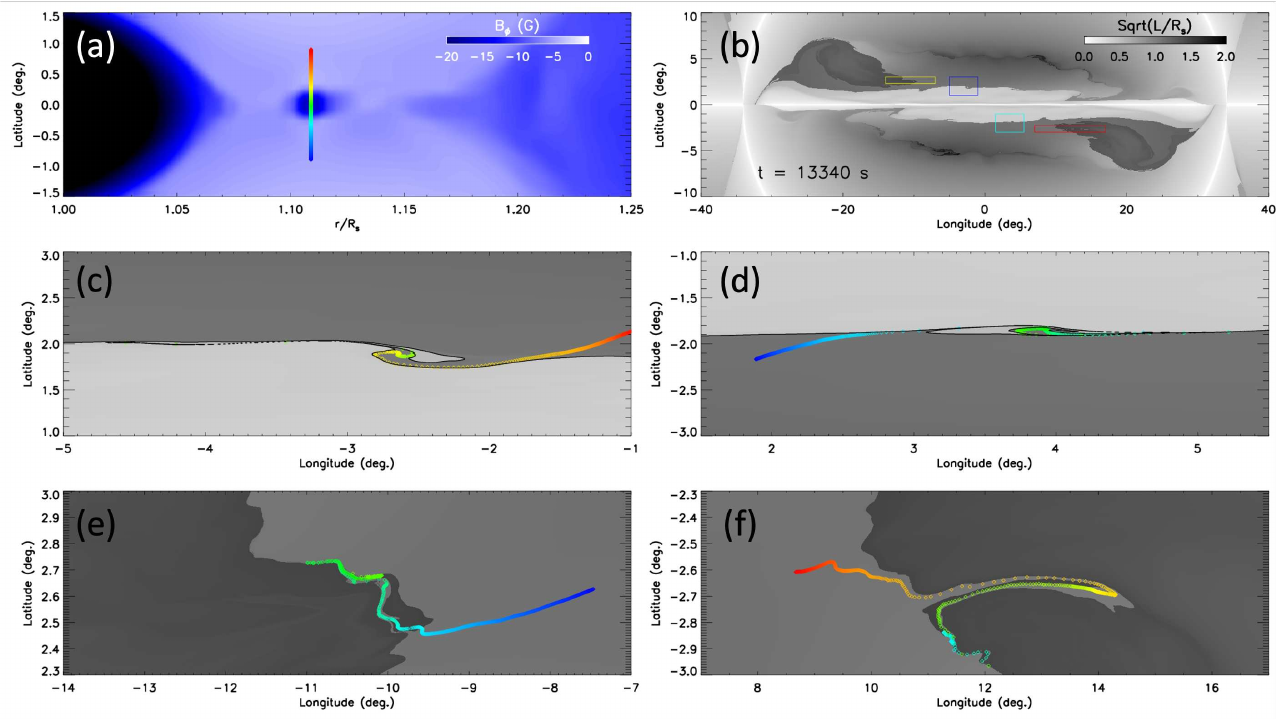}
\caption{
Illustration of the four imprints of an upward-propagating plasmoid on the $L$-map at $t = 13,340$ s. Panel (a) shows an identical slice through the flare current sheet as was shown in Fig.~\ref{fig:merger_up_6frames}. Panel (b) shows the $L$-map of the entire flaring region. Blue and cyan boxes correspond to insets (c, d respectively) showing conjugate footpoints along the parallel section of the ribbons. Yellow and red boxes indicate conjugate patches of the flare ribbon hooks (e, f respectively) that wrap around the flux rope footpoints. An animation of the mapping from the plasmoid cross-section in the corona to the solar surface $L$-map is available, illustrating the temporal evolution of four imprints of the plasmoid during its formation and upward propagation for the interval 13,290 s $\leq$ t $\leq$ 13,400 s at a 1 s cadence (the animation duration is 5 s).
}
\label{fig:merger_up_all_first}
\end{figure}
\begin{figure}[ht!]
\plotone{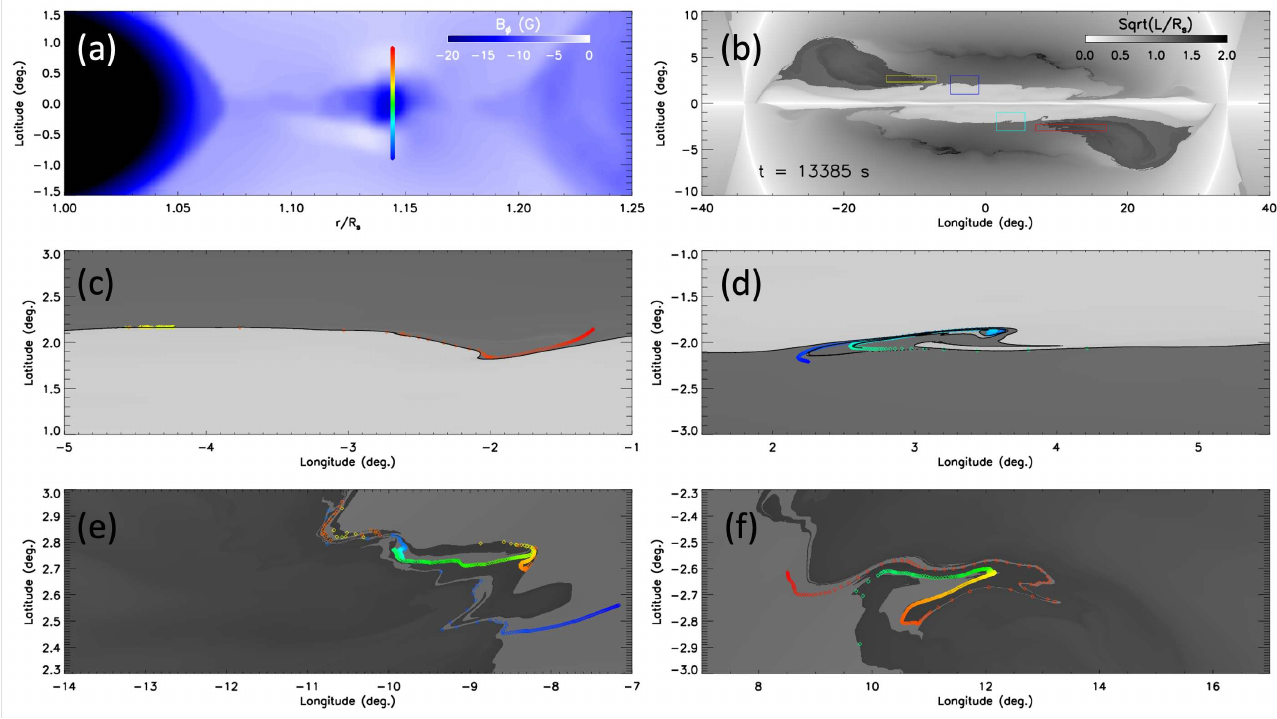}
\caption{
Illustration of the four imprints of an upward-propagating plasmoid on the $L$-map at $t= 13,400$ s. The quantities and regions shown correspond to those depicted in Fig.~\ref{fig:merger_up_all_first}.
}
\label{fig:merger_up_all_last}
\end{figure}

To determine what is responsible for this behavior, we examine the connectivity of this plasmoid within the context of the entire flaring region. We show two different stages of the evolution: $t = 13,340$ s in Fig.~\ref{fig:merger_up_all_first} (when a well-developed spiral is present in Fig.~\ref{fig:merger_up_6frames}) and $t = 13,400$ s in Fig.~\ref{fig:merger_up_all_last} (when no spiral remains). 
In both figures a cross section of the plasmoid is shown (a), illustrating its location and the set of seed points that are traced in both directions to the solar surface. An $L$-map of the entire flaring region is also shown for each case (b), overlaid with rectangles indicating locations of four $L$-map patches shown in (c-f). At $t = 13,340$ s, (Fig.~\ref{fig:merger_up_all_first}), field lines traced from the plasmoid map to \textit{four} distinct imprints: two along the main ribbon (corresponding to the spiral shown in Fig.~\ref{fig:merger_up_6frames} as well as its conjugate), as well as two along the hooks that wrap around the flux rope footprint (Fig.~\ref{fig:merger_up_all_first}e-f). The structure along the hooks is particularly complex; although the bundles of long (dark) and short (light) field lines are wrapped about each other in a similar manner, the overall pattern is not the simple spiral that manifests along the main flare ribbons. Multiple factors are likely to contribute to this complexity, including (a) the location of these features at the extreme ends of the ribbons, (b) multiple wrappings of the interior (dark) field lines about the flux rope axis, and (c) the greater length of these field lines in comparison to those of the flare arcade. Each of these contribute to additional opportunities for distortion of the topological boundaries along this selection of field lines.
\par
Further along in the evolution (Fig.~\ref{fig:merger_up_all_last}) the plasmoid has grown and propagated upward toward the CME flux rope. No spiral feature remains in Fig.~\ref{fig:merger_up_all_last}c, and although significant structure is present in Fig.~\ref{fig:merger_up_all_last}d, this structure is primarily due to the formation of a new plasmoid sunward of the reconnection site ($r \approx 1.09 R_s$ in Fig.~\ref{fig:merger_up_all_last}a). The fine-scale features on the flux rope hooks, on the other hand, have grown significantly in both area and complexity (Fig.~\ref{fig:merger_up_all_last}e-f). The field lines drawn from the plasmoid are closely interwoven with this structure, reinforcing the physical link between the features.
\par
We suggest that these dynamics may be explained by the following sequence: initially, the coronal structure is comprised of intertwined magnetic flux tubes with distinct topologies: flare loop/downward-propagating (Fig.~\ref{fig:plasmoid_types}g) vs. flux rope sheath/upward-propagating (Fig.~\ref{fig:plasmoid_types}h). This configuration could correspond to a small scale flux rope centered on a Hyperbolic Flux Tube (HFT, the intersection of quasi-separatrix layers that represents a three-dimensional generalization of an X-line \citep{titov03a}) as was explored in WP21 (their Fig. 5a), and might in general arise from three-dimensional plasmoid dynamics (e.g., interaction and reconnection of oblique modes on distinct flux surfaces \citep{daughton11a,dahlin17a}). This intertwined structure is transient, however, and eventually the plasmoid evolves into an entirely upward-propagating topology with no remnant imprint on the core/straight sections of the flare ribbons. The long-term dynamics and ribbon imprints of this structure are in line with the analytical model of WP21, who argued that plasmoids ejected upward would have imprints along the flux rope hooks. Nonthermal emission near the footpoints of the CME flux rope have recently been identified in microwave\citep{chen20b} and hard X-ray\citep{stiefel23a} observations. Such observations, though uncommon, could be explained by the release of energetic particles by upward-propagating plasmoids that merge with the outgoing flux rope.
\par
It is apparent that $L$-maps encode rich information regarding plasmoid formation and evolution in the current sheet. The simple case shown in Fig.~\ref{fig:merger_down_6frames} reveals how the formation and filling-in of the structures reflects the generation, propagation, and annihilation of the plasmoids. The example of the upward-propagating plasmoid in \cref{fig:merger_up_6frames,fig:merger_up_all_first,fig:merger_up_all_last} further fleshes out this picture, illustrating not only how such plasmoids imprint primarily on the flux rope hooks, but hinting at complex behavior resulting from the interaction of distinct structures in the current sheet. Poetically, we might declare that flare ribbons tell the story of the plasmoid life cycle: \text{birth, search for direction, maturation, and ultimate fate}.

\section{Synthetic Observations}\label{sec:observations}

In the previous section, we demonstrated how plasmoid dynamics imprint complex, fine-scale features on the solar surface. We now proceed to derive predictions for observable features associated with plasmoids. We first examine coronal signatures for limb observations by constructing synthetic white-light images of the flare current sheet which reveal `blobs' of enhanced density corresponding to plasmoids. We then investigate a simple proxy for flare ribbons on the disk, which we use to illustrate the formation and disappearance of spirals along the ribbon front. For both cases, we present quantitative predictions for the properties of plasmoids and their ribbon imprints.

\subsection{Synthetic White Light Images\label{subsec:whitelight}}

\begin{figure}[ht!]
\plotone{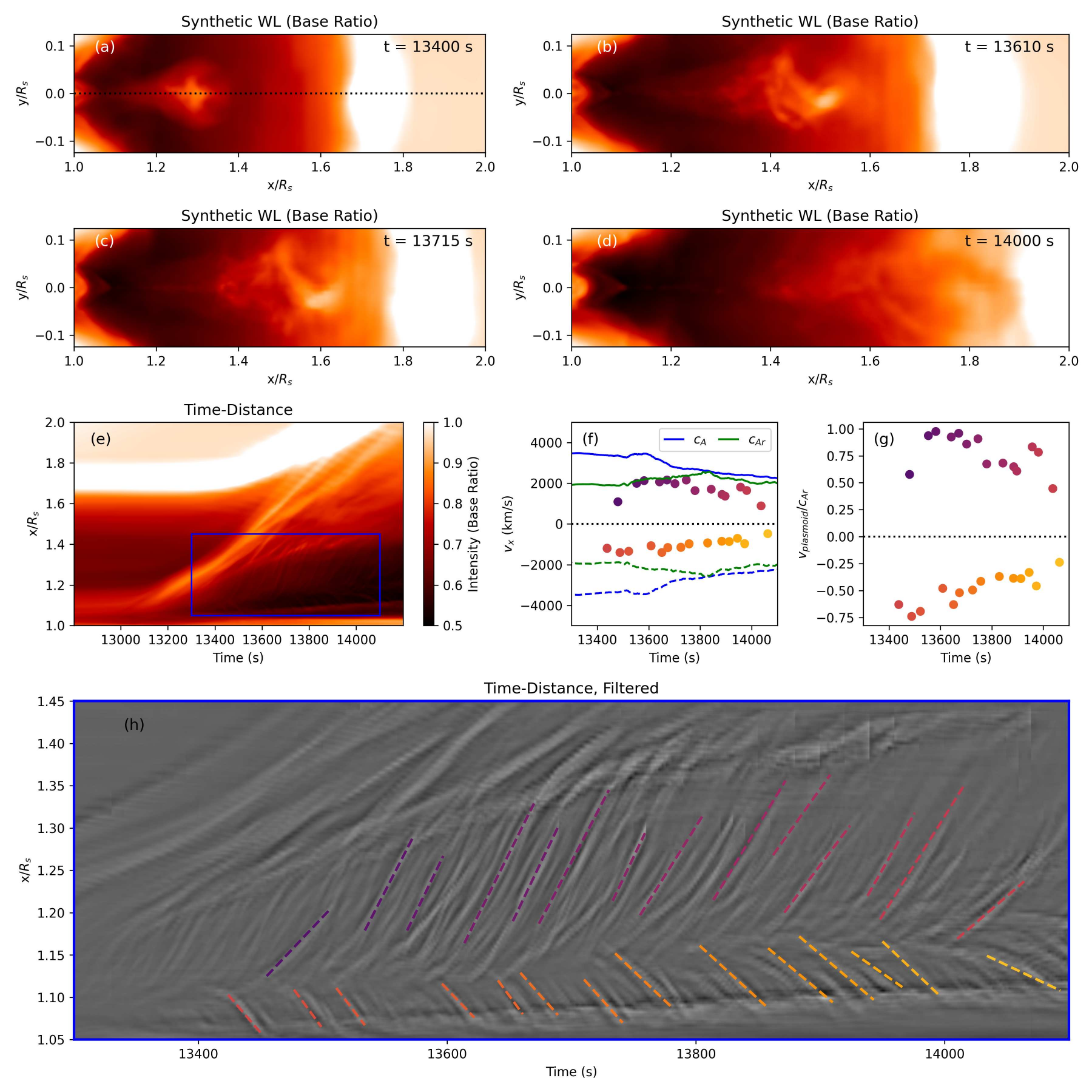}
\caption{Synthetic white light images illustrating plasmoid dynamics in the flare current sheet. (a-d) Snapshots of the flare current sheet at four times, showing the formation and propagation of the CME shock, flux rope core, and flare current sheet containing faint `blobs' corresponding to plasmoids. (e) Time-distance plot drawn from the slit indicated in (a). Faint tracks in the bottom-right (e.g., inside the boxed region) indicate plasmoid trajectories. (f-h) Detailed properties of a selection of plasmoid trajectories. (f) Velocities as a function of time corresponding to the fits illustrated in panel (h), plotted against the upstream Alfv\'en velocity calculated using the total magnetic field ($c_A$) and only the reconnecting component ($c_{Ar}$). (g) Plasmoid velocities normalized to $c_{Ar}$. (h) High-pass filtered version of the boxed region indicated in  (e). Plasmoid tracks are indicated by dashed lines with colors corresponding to the points in (f,g). An animation of the synthetic white-light images is available,  at a $1$~s cadence for $12,800$~s $\leq t \leq 14,198$~s (the animation duration is 23.3 seconds).}
\label{fig:wl_summary}
\end{figure}

Investigations of the dynamics of the low corona during flares typically focus on UV/EUV imaging and spectroscopy, which contain information about plasma temperature, composition, and line-of-sight flows. The detailed physics required for synthesizing such signatures is beyond the scope of our present model. We aim instead to derive a simple tracer for plasmoid dynamics. For this purpose, white-light coronagraph images, which depend only on the electron density (and hence encode information about density structure), are sufficient for identifying plasmoid locations and trajectories.
\par
We use the same method for constructing synthetic white-light images as was employed in previous ARMS investigations \citep{vourlidas13a,lynch16a,lynch21a,lynch25a}. The procedure, which was adapted from the SolarSoft routine \texttt{eltheory.pro} \citep{billings66a,vourlidas06a}, calculates the total brightness I(t) by integrating the line-of-sight white-light intensity associated with Thomson scattering. Base ratio images $I_n(t) = I(t)/I(0)$ are used in order to compensate for the initial background density profile and emphasize small-scale features. 
We chose a viewing perspective such that the center of the flare current sheet is seen edge-on on the west limb. Line-of-sight integration is performed using 256 equally spaced point samples of the plasma density over the range $-2 R_s < l < 2 R_s$, where $l = 0$ corresponds to the plane $\phi = 0$.
\par
A representative selection of synthetic white light images are shown in Figure~\ref{fig:wl_summary}. The formation and eruption of a CME flux rope with filamentary internal structure is evident in Fig.~\ref{fig:wl_summary}a,b. Development and propagation of a CME shock is indicated by the bright region $I_n > 1$ further out from the flux rope. We note that the base ratio in the vicinity of the erupting structure is less than unity due to expansion resulting from the shearing, a feature commonly observed in coronal cavities \citep[e.g.,][and references therein]{parenti14a} associated with filament channels. Faint blobs along the flare current sheet ($\approx$ equator) correspond to plasmoids. These blobs appear as tracks in the time-distance plot (Fig.~\ref{fig:wl_summary}e) constructed from the slit indicated in Fig.~\ref{fig:wl_summary}a. The acceleration of the CME flux rope and associated shock are also apparent in the time-distance figure. 
\par
The properties of the blob trajectories are illustrated in Fig.~\ref{fig:wl_summary}(f-h). We generate a filtered time-distance image (h) by applying a Gaussian filter ($\sigma =$ 3 pixels for both $x, t$) and subtract the result from the original image (e). The result clearly highlights plasmoid tracks. We indicate maximum velocities with dashed lines. A scatter plot of the plasmoid velocities, with color indicating the associated fit in (h), is shown in Fig.~\ref{fig:wl_summary}(f,g). Figure~\ref{fig:wl_summary}f also shows the upstream (i.e, just outside the current layer) evolution of two forms of the Alfv\'en velocity (the conventional Alfv\'en velocity, $c_A$, as well as a velocity calculated only from the reconnection component of the magnetic field ($c_{Ar} = (B_{rec}/B)c_A$). The latter represents a stronger theoretical constraint on the velocity that results from the fact that only the reconnecting component of the magnetic field contributes to the tension force that drives the outflow. Indeed, we see that the plasmoid velocities do not exceed this limit (panel g), and furthermore that there is an overall trend of diminishing plasmoid velocity. We also note that the mean upward velocity ($\left<v_{up}\right> \approx 1600$ km s$^{-1}$ is significantly larger than the mean downward velocity $\left<v_{down}\right> \approx -1000$ km s$^{-1}$. Two related factors are likely at play: (1) downward propagating plasmoids soon encounter the stationary flare arcade hindering acceleration, whereas the upward propagating plasmoids must `catch up` to the ejected CME flux rope; (2) the primary reconnection X-line is rising, resulting in a net upward shift in the velocities measured in the simulation frame.
\par
Numerous observations have shown similar localized upflows and downflows \citep{takasao12a,takasao16a,liu13a_w,kumar13a,kumar19a,kumar23a,kumar25a,cheng18a,yu20a} with typical velocities ranging from a few $100 \text{km s}^{-1}$ up to $> 1000$ km s$^{-1}$ \citep{liu13a_w,cheng18a}. In several of these examples, the mean downflow velocity is substantially smaller the mean upflow velocity \citep{liu13a_w,cheng18a,kumar19a,kumar25a}, in line with the pattern identified in the simulation. While the velocities indicated in Fig.~\ref{fig:wl_summary} are on the high end of what is observed, the magnitude scales with the Alfv\'en velocity, which can vary widely in the corona and as a function of the guide field. Other important observational factors may include limited temporal resolution that can hinder identification of fast-moving features with $v \gtrsim 1000 $km s$^{-1}$ \citep[e.g., as noted by][]{lorincik24a} and projection effects that reduce the apparent velocity in the plane-of-sky.
\par
The velocity of these structures is also worth comparing with Supra-Arcade Downflows (SADs) \citep{mckenzie00a,savage12a}, underdense finger-like features that are observed to descend toward the flare arcade. SADs have been measured to have typical velocities $\sim 50 - 500 $km s$^{-1}$, which has been noted to be much smaller than what is expected for the coronal Alfv\'en velocity. The contrast with the simulation results may therefore reinforce the interpretation that SADs do not correspond to reconnected loops, but rather are manifestations of instabilities in the turbulent interface region between the flare loops and the reconnection outflow \citep[such as would occur below a flare termination shock][]{shen22a}. In this discussion, it is also worth noting that the density enhancements highlighted in Fig.~\ref{fig:wl_summary} decelerate rather suddenly upon reaching either end of the flare current sheet (acceleration and deceleration have been noted in many observations of both bright and dim features), and it may not be obvious from a given observation whether a propagating feature is better identified with a structure within or downstream of the flare current sheet. 
\par

\subsection{Synthetic Flare Ribbons \label{subsec:ribbons}}

\begin{figure}[ht!]
\plotone{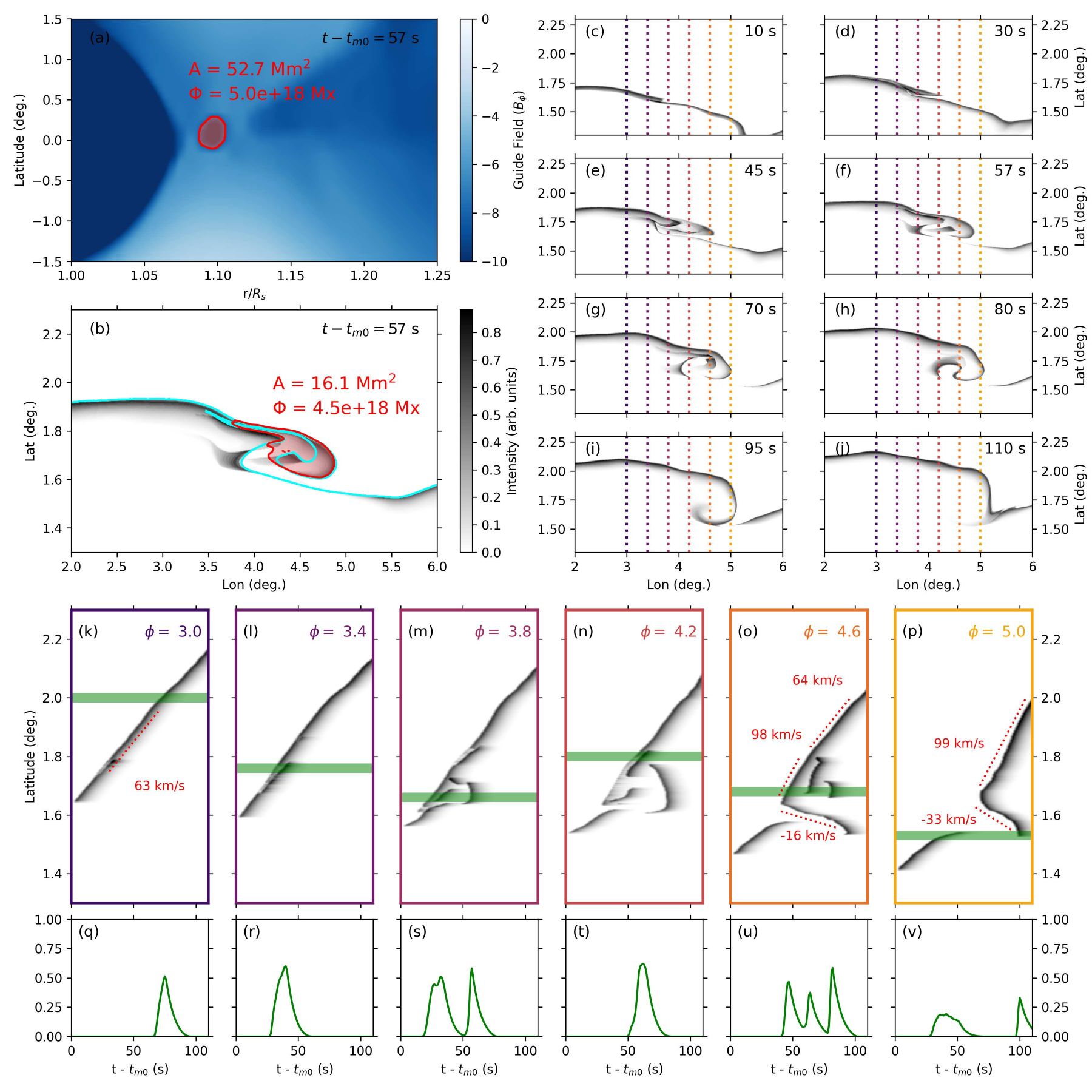}
\caption{Synthetic ribbon proxy for the evolution of the downward-propagating plasmoid (Fig.~\ref{fig:merger_down_6frames}. (a) Plasmoid cross section at the time of its maximum area. Blue shading indicates the guide field $B_\phi$ and red indicates a selected flux bundle that was traced to the solar surface. (b) Flare ribbon proxy (grayscale) at the solar surface superposed with a mapping (red) of the plasmoid flux indicated in panel (a) and an $L$-map contour (cyan) corresponding to $L = 0.6 R_s$. (c-j) Snapshots of the synthetic flare ribbon patch (indicated times are measured relative to $t_{m0} = 13,290$ s). Time-distance plots generated at the locations of the slits indicated in (c-j) are shown in (k-p); the longitude coordinate ($\phi$) is indicated for each panel. Light curves corresponding to the latitude-averages of the regions indicated by the green slits in (k-p) are shown in (q-v). An animation is available showing the evolution of the synthetic flare ribbons at a $1$~s cadence for $13,290$~s $\leq t \leq 13,400$~s (the animation duration is 4.6 seconds). The intensity is represented with a black-red-yellow-white colormap to better evoke solar imagery.}
\label{fig:rib_summary}
\end{figure}

We next derive a synthetic proxy for the flare ribbons. As noted previously, full forward modeling of flare ribbons would require detailed treatment of the generation, propagation, and deposition of non-thermal particle energy. These processes remain an active area of research in 1-D and 2-D numerical modeling (e.g., loop models such as \texttt{RADYN} \citep{allred05a} and hybrid particle-MHD codes such as \texttt{kglobal} \citep{drake19a,arnold21a}), and currently represent a prohibitive expense in 3-D MHD modeling. We instead aim to construct tracers of the flare ribbon front in order to identify morphological features that may be compared to observations (a similar strategy to our approach for coronal images), which we will likewise show is sufficient to reveal important quantitative information about the link between the ribbon imprints and the current sheet plasmoids.
\par
For simplicity, we assume that the onset of brightening at a given location of the solar surface represents an instantaneous response to reconnection along the associated field line. Flare reconnection events are identified using a sequence of $L$-maps (a value of 1 is assigned for a given footpoint at time $t = t_{rec}$ if $L(t)-L(t-\Delta t) < -L_{rec}$, and 0 otherwise). The threshold $L_{rec} = 0.2 R_s$ was determined by inspection of the data ($\Delta L$ never takes values between -0.22 and -0.11 for this dataset, so this choice cleanly selects field-line shortenings that are due to reconnection). The result is then convolved with an ad-hoc light curve consisting of a rapid rise to peak intensity followed by an exponential decay (see for example Fig.~\ref{fig:rib_summary}q) according to the following expression:
\begin{equation}
    I_{lc}(t') =
    \begin{cases}
        t'/t_{rise},  & 0 \leq t' \leq t_{rise}  \\
        A\exp\left[-\frac{t'-t_{rise}}{\alpha t_{decay}}\right]-B, & 0 \leq t' - t_{rise} \leq t_{decay} \\
        0 & \text{otherwise}
    \end{cases}       
\end{equation}
where $t' = t - t_{rec}$ , $\alpha$ parametrizes the exponential decay, and $A$ and $B$ are chosen so that $I_{lc}(t_{rise}) = 1$ and $I_{lc}(t_{rise}+t_{decay}) = 0$. We chose $t_{rise} = 2$ s, $t_{decay} = 20$ s, and $\alpha = 4 \ln 2$. These values are somewhat small compared to what is typically reported in observations. However, rise times $\sim 10 - 20$ s have been observed with IRIS \citep{graham20a}, and one study identified rise times as small as $\sim 4$ s with exponential decays $\sim 10$ s \citep{penn16a}. New short-exposure SolO EUI/FSI observations may shed further light on the time scales associated with flare ribbons \citep{collier24a}. Nevertheless, the selected parameters are not reflective a particular observation, but rather were chosen in order to adequately illustrate the evolution in this short ($110$ s) high-cadence interval of simulation data.
\par
The resulting synthetic ribbon proxy for the downward-propagating plasmoid (Fig.~\ref{fig:merger_down_6frames}) is illustrated in Fig.~\ref{fig:rib_summary}.
In order to examine the quantitative properties (magnetic flux, area) of a plasmoid and its imprint, we mapped the plasmoid from the corona to the surface using the following procedure. We first identify the cross-section of the plasmoid in the corona (for the snapshot in Fig.~\ref{fig:rib_summary}, we select the region $B_\phi < -8$ G in the vicinity of a minimum in the `in-plane' magnetic field). A grid of field lines was then traced from this region to the solar surface. We use the resulting footpoint locations to construct a Gaussian kernel density estimation for the mapping of the plasmoid magnetic flux using the \texttt{SciPy gaussian\textunderscore kde} function \citep{virtanen20a}. In order to minimize the impact of outliers of the field-line mapping, we selected a contour of this density that enclosed 90\% of the magnetic flux. This mapping is illustrated in Fig.~\ref{fig:rib_summary}(a-b).
\par
The cross-section of the plasmoid in the corona is indicated in Fig.~\ref{fig:rib_summary}a, which shows the plasmoid mask (red) superposed on contours of the magnetic guide field ($B_\phi$). The magnetic flux enclosed by the red contour maps to a reversed `sigma` shaped patch on the solar surface (b) that is co-located with a `spiral' along a patch of the proxy ribbon (grayscale) and an $L$-map contour (cyan) that indicates the boundary between reconnected and unreconnected flux. The time-evolution of the synthetic ribbon structure is shown in Fig.~\ref{fig:rib_summary}(c-j) and exhibits a spiral feature similar to what was illustrated in the corresponding \textit{L}-maps (Fig.~\ref{fig:merger_down_6frames}).
\par
Time-distance plots (Fig.~\ref{fig:rib_summary}k-p) were extracted from the slits indicated in the snapshots, and `light curves' (Fig~\ref{fig:rib_summary}q-v) constructed from latitude averages of the green slits are shown in (k-p). These figures illustrate several distinct features that emerge from this simple ribbon proxy. A given 1-D slit can show a variety of complex behavior, including brightenings ahead of the primary ribbon (n, o) and reversals of the apparent ribbon propagation direction (o, p). Selected apparent velocities, indicated with dashed lines, show that the apparent velocity for a given slit may be significantly larger (e.g., o, p) than the typical mean velocity (k). The light curves (q-v) also exhibit a variety of behaviors including (1) multiple-peaks (s, u, v) and (2) longer-duration features (s, v). We emphasize that the fine structure and complex dynamics arise from a simple ad-hoc model for the flare reconnection and resulting footpoint emission, which is nevertheless sufficient to mimic the spiral morphology observed in observations.
\par
The spiral shown in \Cref{fig:rib_summary} is roughly elliptical, with a major axis measuring $\lesssim 10 $Mm. This is comparable to the largest structures identified in IRIS data (e.g., Fig. 1 in WP21). However, these were among the the largest such features observed, and are not typical of flares, and many observations have revealed features as small as $~\sim 100$--$200$~km \citep[e.g.][]{jing16a,collier24a,faber25a}. To examine how the spiral properties in our idealized simulation model might be applied to observations, we can derive a simple scaling relation that illustrates the various impacts of model parameters (e.g., characteristic length scale and magnetic field strength). Magnetic flux ($\Phi$) must be conserved along the flux tube linking the plasmoid to the solar surface, i.e., $\Phi_{plasmoid} = \left<B_{plasmoid}\right>A_{plasmoid} = \left<B_{spiral}\right>A_{spiral}$, 
where $<>$ indicate spatial averages for a given feature (spiral or plasmoid). For the plasmoid shown in Fig.~\ref{fig:rib_summary}(a,b), the magnetic field is weaker in the corona $\left<B_{plasmoid}\right> \approx -9.5$ G $< \left<B_{spiral}\right> \approx -28$ G, whereas
$A_{plasmoid} \approx \SI{52.7}{\square\mega\meter} > A_{spiral} \approx \SI{16.1}{\square\mega\meter}$, illustrating the inverse relationship between the area and magnetic field strength.
\par

In order to apply this expression to observations, we can rewrite this relation in more tangible form by dropping the spatial averages and inserting typical values for the solar atmosphere to obtain the following expression for the area of a spiral:

\begin{equation}
A_{spiral} \sim \si{\square\mega\meter}
\left(\frac{A_{plasmoid}}{\SI{10}{\square\mega\meter}}\right)
\left(\frac{B_{plasmoid}}{\SI{100}{\gauss}}\right)
\left(\frac{\SI{1000}{\gauss}}{B_{spiral}}\right)
\label{eqn:spiral_scaling}
\end{equation}

We may use this scaling relation to identify several reasons why a typical flare would feature smaller spiral structures than in our idealized simulations.
First, the inner boundary of the model represents the base of the corona, corresponding to a magnetic field strength that may be significantly weaker than in the photosphere/chromosphere where flare ribbons form. This is only partially offset by the model magnetic field strength in the corona being somewhat weaker ($~10$ G) than is typical in flares, and the net effect is that the expansion of the magnetic field ($B_{spiral}/B_{corona}$) is greater on the sun than in the model. A second issue is that the model generates larger plasmoids that are typically observed in flares, which is a direct consequence of the spatial scale employed for the idealized active region. This spatial scale is the primary factor controlling the length of the current sheet, which in turn constrains the maximum plasmoid size (in the model, $L_{plasmoid} \lesssim 0.1 L_{flare\_cs}$; see for example Fig.~\ref{fig:rib_summary}a). The MHD model also only resolves the upper end of the the range of plasmoid sizes, which would be expected to extend to kinetic scales. Hence, it is not surprising that the plasmoid shown in Fig.~\ref{fig:rib_summary} has a diameter $\sim$ 10 Mm is significantly larger than has usually been measured in observations, where `blobs' typically span a few Mm (e.g., \textcite{takasao12a,kumar19a}). 
\par
Given these factors, we should not expect flare observations to exhibit features quite so large as the structures identified in the simulation. Nevertheless, the scaling relation in Eq.~\ref{eqn:spiral_scaling} would suggest that `spirals' might be expected to have scales $\sim \si{\square\mega\meter}$, which is more typical of fine structures in observations. Returning to the striking `breaking waves' from the Sept. 2014 IRIS X-flare, we point out that the large spatial scale associated with these features is likely due to the weak magnetic fields in these regions which would cause the plasmoid footpoints to spread over a larger region ($A_{spiral} \propto 1/B_{spiral})$.
\par
In this section, we have presented synthetic observations that represent tracers for the qualitative morphological signatures of plasmoids in the flare current sheet and in the ribbons. Full forward modeling (detailed thermodynamics, radiative transfer, and treatment of the lower solar atmosphere) would be required for quantitative prediction of emission signatures (e.g., Doppler shifts). Nevertheless, these synthetic observations contain quantitative information about the dynamics that might be identified from observations. Tracers are indeed sufficient for a variety of routine observational calculations such as magnetic flux reconnection rates  \citep{qiu04a} or velocities and acceleration of upflows/downflows along the current sheet \citep{liu13a_w,kumar13a,yu20a,kumar25a}. Based on a simple mapping of plasmoid flux to the solar surface, we estimate that the spatial scale associated with ribbon spirals should be $\lesssim$~Mm for typical flares.

\section{Discussion}\label{sec:discussion}

In the investigation described above we have shown how the fine structure observed in flare ribbons reveals the structure and dynamics of the magnetic reconnection in a flare current sheet, in particular the role of plasmoids. These results represent the first such demonstration within a three-dimensional eruptive flare model that self-consistently captures the formation and evolution of 3D plasmoids. Importantly, we have shown how the spirals imprinted by the ribbons encode information about the plasmoid `life cycle', i.e., formation, propagation, and destruction by reconnection either with the flare arcade or CME flux rope.

By design, the eruptive flare model explored in this study was highly idealized, with a high degree of symmetry and a smooth initial magnetic field configuration in order to identify as clearly as possible the impact of plasmoids on flare ribbon structure. In the observations, however, a variety of other factors are likely to play a significant role in the structuring of flare ribbons, which include for example the convolution with the local magnetic field distribution of the chromosphere/photosphere (which is itself highly structured) or variable heating and cooling times. Our model also does not exclude the roles of other physical processes (such as slipping reconnection or Kelvin-Helmholtz instabilities). 
\par
Regardless,  well-established theory and modeling has found that reconnection in systems such as the solar corona (which has a vast separation between the global and dissipative scales) is expected to be mediated by plasmoids\citep{shibata01a,ji11a}. It would in fact be surprising if plasmoids did \textit{not} play a significant role in the reconnection process. Given the likely importance of plasmoids to the flare dynamics, we therefore suggest that our model represents an important lens through which to interpret and deconvolve the various factors involved in the fine structure of flare ribbons.

Encouragingly, our study confirms several key results previously demonstrated with the static analytic model explored by \citep{wyper21a} (WP21). These include:
\begin{itemize} 
\item The chirality of the ribbon spirals is in the same sense as the flux rope footprint (compare Fig.~\ref{fig:flare_context} to WP21 Fig. 13).
\item The formation and propagation of the plasmoid toward the flare arcade manifests as the `filling in' of the spirals with reconnected flux (WP21 Fig. 9).
\item The spiral patterns drift parallel to the ribbons (compare Fig.~\ref{fig:rib_summary} to the prediction in WP21 Fig. 10).
\end{itemize}
Furthermore, the following results extend the conclusions drawn from the analytic model:
\begin{itemize}
\item The present model reveals complex topologies that imprint transient spirals at all four relevant footpoints of the plasmoids (along each of the conjugate straight and hooked sections of the ribbon).
\item The detachment of a plasmoid from a ribbon patch (Fig.~\ref{fig:merger_up_6frames}) manifests as a`retraction'/`unwinding' of the spiral, rather than as a `filling-in'.
\item The formation of multiple plasmoids and the interactions of their topological imprints leads to nested structuring of the ribbons (a result hinted at in Fig.~\ref{fig:merger_up_all_last} and anticipated by the 3-D null point study of \citet{wyper14a}).
\end{itemize}
\par
The results of the present study spur a number of intriguing questions for further investigation, which include but are not limited to the following:
\begin{itemize}
\item Can flare ribbon fine structure be used to infer plasmoid statistics (for example distinguishing between the size distribution predicted by reconnection models \citep{fermo10a,huang12a})? 
\item What are the implications of three-dimensional plasmoid structure on particle acceleration and transport \citep[e.g.][]{dahlin15a,dahlin17a,li19a,dahlin20a}?
\item What is the role of upward-propagating plasmoids in the flare particle acceleration process and where do they go, given that weak non-thermal emission at the flux rope footpoints and energetic electrons fluxes in the solar wind\citep[e.g.,][]{krucker07a} indicate much smaller populations than are present in the flare ribbons?
\item How does the turbulent interaction of many plasmoids manifest in the topological imprints, and can this be observed?
\item How does plasmoid structure (and its corresponding imprint) vary with the properties of the flare (e.g., confined vs. eruptive; strong vs. weak guide field)?
\end{itemize}

A topic of interest in broader heliophysics is the possible relationship between flare ribbon features and the auroral structure observed at the Earth. In both contexts, a remote energy release process drives energetic particles, which precipitate into a high density plasma interface triggering emission that reveals the magnetic imprint of the source. Similar to the ribbon features, the aurora are highly dynamic and often characterized complex fine-scale features such as spirals, folds, and curls \citep{hallinan76a}. Notably, the spirals exhibit a consistent chirality, with a rotational sense that is clockwise when viewed anti-parallel to the magnetic field \citep{davis76a,partamies01a}. One proposed explanation is reconnection (tearing) in the field-aligned current sheets of the auroral acceleration region \citep{huang22a_k}, which has intriguing similarities to the flare ribbon scenario, but also distinct differences (the more direct analogy for the flare scenario would be reconnection in the magnetotail, rather than in the field-aligned currents). Further investigation is warranted to clarify to what degree these phenomena involve the same physical origin.

We have presented in detail a simple yet powerful tool for analyzing the topological evolution of eruptive flares, the field-line length map (\textit{$L$-map}). Abrupt changes in the field-line length may be transparently identified as resulting from reconnection events, making it especially useful for identifying model proxies of flare ribbons and tracking their evolution \citep[as previously demonstrated in][]{dahlin19a,lynch19a,lynch21a,dahlin22a}. Furthermore, the lengthening of field lines is an indicator of regions where coronal dimming might be expected, for example near the flux rope footpoints).
As with Q-maps, the $L$-map encodes similar information about the global connectivity of the system and the distinct topological domains (indeed, Q-maps were employed in WP21 to show how plasmoids distort QSLs associated with flare ribbons). However, while Q-maps are excellent tools in particular for identifying finite \textit{gradients} in the connectivity that may be subject to reconnection, $L$-maps more clearly reveal the relationships between distinct neighboring topologies (compare Fig.~\ref{fig:ribbonproxies}a,c) and the nature of the reconnection that occurs when the connectivity changes. While $L$-maps are particularly useful for analysis of 3D numerical models, where the connectivities are precisely known, we suggest that $L$-maps may also be helpful for interpreting coronal magnetic field extrapolations.
\par
Finally, our results provide key predictions for observations of flare ribbons. The instantaneous ``spirals" that are evident in the $L$-map figures above have already been seen in high-resolution UV and EUV observations. We expect that instruments such as EUVI on Solar Orbiter with its ultra-high spatial and temporal resolution will show many more such structures, perhaps allowing us to obtain quantitative estimates of the numbers and scales of plasmoids resulting from flare reconnection. Important predictions from our results are the single-point light curves shown in \Cref{fig:rib_summary}, in particular, the ones with multiple peaks. It should be possible measure such light curves in the data. If so, they could provide invaluable new information on the time scales for plasmoid formation and propagation in flares, and on the dynamics of solar reconnection in general.

\begin{acknowledgments}

JTD was supported by NASA H-LWS grant 80NSSC21K0816, H-ECIP grant 80NSSC23K1065 and H-TMS grant 80NSSC24K0608. SKA was supported in part by a NASA Living With a Star grant and a NSF SHINE grant to the University of Michigan.
PFW was supported by an STFC (UK) consortium grant ST/W00108X/1 and a Leverhulme Trust Research Project grant. JQ was supported by the NASA HGI grant 80NSSC23K0414, and the NSF grants AST-2407849 and AGS-2401228.
We acknowledge the use of Peter Wyper's topology tools \citep{wyper16a,wyper16b} and Ben Lynch's synthetic white light imaging tools \citep{lynch25a}. JTD acknowledges Masha Kazachenko for coining the term `$L$-map`. Computational resources supporting this work were provided by the NASA High-End Computing (HEC) Program through the NASA Center for Climate Simulation (NCCS) at Goddard Space Flight Center and the Texas Advanced Computing Center (TACC) at The University of Texas at Austin.
\end{acknowledgments}

\providecommand{\noopsort}[1]{}\providecommand{\singleletter}[1]{#1}%

\bibliographystyle{aasjournal}

\appendix
Given the importance of the $L$-map to the analysis presented in this paper, we present additional discussion for its practical use here. Field lines are constructed using full magnetic field data output from ARMS using topology tools developed by Peter Wyper \citep{wyper16a,wyper16b,wyper24a}, which employs RK4 integration to solve the field line equation: $d\mathbf{r}/ds = \mathbf{b(r)}$, where \textbf{r} is the position vector, $ds$ is the step length, and $\mathbf{b(r)} = \mathbf{B(r)}/|\mathbf{B(r)}|$. 

The field-line length $L = \int_{s0}^{s1} ds$ is the length of the integrated curve from its source point to where it meets a boundary in the domain. In the present case, all field lines in the region of interest close to the solar surface. To generate a time-evolving `map', we select a grid of uniformly spaced seed points at $r = R_s$ from which we trace field lines at a specified temporal cadence. Table~\ref{tab:sim_params} summarizes the parameters for each of the $L$-maps shown in this article.

\begin{table}[htbp]
\centering
\caption{L-Map Parameters}
\label{tab:sim_params}
\begin{tabular}{lccccc}
\hline
\textbf{Name} & \textbf{Lon. Pts.} & \textbf{Lat. Pts.} & \textbf{Lon. Range} & \textbf{Lat. Range} & \textbf{Cadence} \\
\hline
Fig.~\ref{fig:lmap_explainer} & \(2000\) & \(500\) & \([-40~,~40]\) & \([-10~,~10]\) & 10 s \\
Fig.~\ref{fig:ribbon_proxy_4frames}b & \(2000\) & \(500\) & \([-10~,~10]\) & \([0~,~5]\) & 10 s \\
Fig.~\ref{fig:merger_down_6frames} & \(2000\) & \(1000\) & \([2~,~6]\) & \([1~,~3]\) & 1 s \\
Fig.~\ref{fig:merger_up_all_first}c & \(2000\) & \(1000\) & \([-5.5~,~-1.5]\) & \([1~,~3]\) & 1 s \\
Fig.~\ref{fig:merger_up_all_first}d & \(2000\) & \(1000\) & \([1.5~,~5.5]\) & \([-3~,~-1]\) & 1 s \\
Fig.~\ref{fig:merger_up_all_first}e & \(3500\) & \(350\) & \([-14~,~-7]\) & \([2.3~,~3.0]\) & 1 s \\
Fig.~\ref{fig:merger_up_all_first}f & \(5000\) & \(350\) & \([7~,~17]\) & \([3.0~,~-2.3]\) & 1 s \\
\hline
\end{tabular}
\end{table}



\end{document}